\documentstyle[aps,amsmath,twocolumn,graphicx]{revtex}
\begin{document}
\wideabs{
\title{A new relation between pressure and fractional flow in two-phase flow in porous media}
\draft
\author{Henning Arendt Knudsen\cite{emailhak} and Alex Hansen\cite{emailah}}
\address{Department of Physics, Norwegian University of Science and Technology, NTNU, NO-7491 Trondheim, Norway}
\date{\today}
\maketitle
\begin{abstract}
  We study average flow properties in porous media using a two-dimensional network simulator. It models the dynamics of two-phase immiscible bulk flow where film flow can be neglected. The boundary conditions are biperiodic which provide a means of studying steady state flow where complex bubble dynamics dominate the flow picture. We find fractional flow curves and corresponding pressure curves for different capillary numbers. In particular, we study the case of the two phases having equal viscosity. In this case we find that the derivative of the fractional flow with respect to saturation is related to the global pressure drop. This result can also be expressed in terms of relative permeabilities or mobilities, resulting in an equation tying together the mobilities of the two phases.
\end{abstract}
\pacs{PACS number(s): 47.55.Mh}
}

\section{Introduction}
Two-phase displacements in porous media have during the past decades been studied by experimental work\cite{CW85,MFJ85,LTZ88}, numerical simulations\cite{KL85,LTZ88,BK90,R90,CP91,CP96}, statistical models\cite{WS81,WW83,P84}, and differential equations\cite{BC98}. The work in this field has been motivated by applications in oil recovery, and also hydrology. The complex nature of transport processes in porous media makes it worthwhile to approach the problem in many ways.

In this paper we present simulation results based on a network simulator of a porous medium\cite{KAH00}. The model is on the pore level scale in the sense that each pore is represented in the network. The generic building blocks are tubes which constitute all inter-pore connections and which hold the pore-space volume. As will become clear in section \ref{sec:model}, each tube allow for a maximum of two fluid-fluid interfaces. This captures essential properties of porous media throats. These aspects set the level of coarse graining of the system. There has been done numerous work on lower length scales and mesoscale methods, such as lattice gas methods\cite{RK88,R90}. An overview over some different scale models can be found in\cite{S95}. On the other hand, compared to large scale reservoir simulations\cite{P77}, our work is detailed and it may provide average properties which can be used as input parameters in work on larger scales. Currently, results are mainly generic, and more detailed work is required to produce results which are valid for specific porous media. In particular, simulations in 3D are required.

In general, the model can be used for 3D porous media and irregular node positions. This follows from the fact that each inter-pore throat of the real porous medium is replaced by a tube in our model. However, in this paper we restrict ourselves to the study of flow in 2D using a regular network with random properties. Previous use of the model in 2D is simulation of drainage\cite{AMHB98,AMH98}. The previous simulations have successfully reproduced the temporal evolution of the three regimes in drainage: viscous fingering, capillary fingering, and stable displacement. The major important innovation in the present work is the change of boundary conditions. Before there was an inlet and an outlet between which an invasion process was simulated. Now biperiodic boundary conditions are used, which makes the system closed, and the system is run for a longer time. This gives average statistical flow properties in a new way, and it is the biperiodic boundary conditions which are the major difference compared to other somewhat similar network models\cite{CP91,CP96}, see Sec. \ref{sec:model}.

The nature of the simulations is so that the systems approach a steady state, in which the two phases flow in a bubble mixture. The notions of drainage and imbibition are not adequate to describe the flow. On pore level there is a continuous process of breaking up bubbles and merging together bubbles. Experimental work showing similar complex bubble dynamics have been done in two-dimensional etched glass networks\cite{AP95a,AP95b,AP99}.

The basic results which are presented are fractional flow curves of the nonwetting phase as a function of nonwetting saturation. Also we present the corresponding global pressure drops which are applied. This is done for six different capillary numbers. The fractional flow and pressure drop can be transformed into the terminology of relative permeabilities and mobilities.

Our simulations show that for the case where the two phases have equal viscosity, there is a relation between fractional flow and pressure. This relation can be expressed in the form of a first order differential equation; the derivative of the fractional flow with respect to saturation is connected to the global pressure drop, also as a function of saturation.

The paper is organized as follows. Section \ref{sec:model} provides a brief description of the model. Section \ref{sec:simulations} concerns simulation results, starting out with describing the nature of the simulations. Thereafter \ref{sec:fraccurves} deals with fractional flow and pressure curves in general, \ref{sec:cadep} discuss how these curves depend on the capillary number, and \ref{sec:presanddfds} establish the relationship between the derivative of the fractional flow and the the global pressure drop. Finally, in \ref{sec:discussion} we make concluding remarks.

\section{Model}
\label{sec:model}
The basis for the present work is a network model which represents the two-dimensional porous medium. The network model is based upon the one presented by Aker \emph{et al.}\cite{AMHB98,AMH98}. The geometric representation of the porous medium and the possible spatial fluid distribution is essentially unchanged. Some changes have been made though. They concern the boundary conditions and the detailed rules for motion of interfaces. All the details concerning these points have been presented in a previous piece of work\cite{KAH00}. Therefore we will restrict ourselves here to an essential r{\'e}sum{\'e} of the physically important aspects and omit the computational details.

The porous medium is represented by a square lattice of cylindrical tubes, tilted $45^{\circ}$ with respect to the overall flow direction. The volume of throats and pores is contained in these tubes. The points where four tubes meet are referred to as nodes. Randomness is incorporated in the system by allowing the position of each node to be randomly chosen within the interval plus minus thirty percent of the lattice constant away from its respective lattice point. From these positions the distance $d_{ij}$ between connected nodes $i$ and $j$ is calculated for all $i$ and $j$. Further, the average radius of each tube is chosen by random in the interval $(0.1d,0.1d+0.3d_{ij})$, where $d$ is the average distance between nodes, i.e. the lattice constant. Clearly, other topologies in 2D (or in 3D) can be chosen. However, as a first approach to the study of the relationship between other variables it is necessary to work with one fixed topology. We will address the topology question again in the discussion. There current restriction to 2D is also very convenient because 3D systems are CPU time demanding.

We consider two fluids within this system of tubes. They are separated by a set of interfaces, {\sl menisci}, in the tubes. We do not allow for film flow. Motion of the fluid during a simulation is represented by the motion of the menisci. In each tube we allow zero, one or two menisci. If at any instant the evolution of the system generates a third meniscus in one tube, then the three menisci are collapsed into one. The position of this new meniscus is the one which preserves the volumes of the two fluids. This upper limit of two menisci in each tube sets the resolution of the fluid distribution.

With respect to permeability the tubes are treated as cylindrical, but with respect to capillary pressure they are hour-glass shaped. This means that over each of the menisci in the system there is a capillary pressure which varies with the menisci's position in the tubes. The formula for the capillary pressure is\begin{equation}
 p_{\rm c} = \frac{2\gamma}{r} (1-\cos{(2\pi x)})
\label{eq:capilpres}
\end{equation}
which is a modified form of the Young-Laplace\footnote{Young-Laplace law is sometimes referred to as Laplace's equation\cite{D92}.} law\cite{D92,AMHB98}. Here, $r$ is the radius of the tube, and $\gamma$ is the inter-facial tension between the fluids. Further, $x$ is the position of the meniscus in the tube, running from zero to one.

The volumetric flux through one single tube is given by the Washburn equation for capillary flow \cite{W21};
\begin{equation}
\label{eq:wash}
  q = - \frac{k}{\mu_{\rm eff}}\cdot\frac{\pi r^2}{d} \left( \Delta p - p_{\rm c}\right) ,
\end{equation}
where $p_{\rm c}$ is given by Eq. (\ref{eq:capilpres}). Considering the tubes as cylindrical with radius $r$, the fraction $\pi r^2 / d$ is the cross-sectional area divided by the length of the tube. The permeability of the tube is $k=r^2/8$, which is known from Hagen-Poiseulle flow. When two fluids are present in a tube, an effective viscosity $\mu_{\rm eff}$ is used. The viscosities of the two fluids are weighted according to their volume fraction within the tube at the beginning of the time step to give $\mu_{\rm eff}$. $\Delta p$ is the pressure difference between the ends of the tube.

In a network of tubes, the net flux passing through the nodes must be zero. There is assigned a pressure to each node. Using the Washburn equation (\ref{eq:wash}) for the flow in a single tube, the flux conservation gives a large system of linear equations in the pressure variables. The equations must be solved using the desired boundary conditions which we will describe shortly.

The traditional use of network simulators has been the study of invasion processes. Usually the first row of nodes is an inlet row where all the nodes have the same fixed pressure. Similarly the last row works as an outlet row, typically having all nodes fixed at zero pressure. For a given time step the pressures in these two rows are fixed constraints. The pressures in all the other nodes are free variables that are solved for.

Once the pressure field is known, Equation (\ref{eq:wash}) gives the flux in each tube. In turn this information is used to forward integrate the system one time step, using the explicit Euler scheme. In practice, forward integration means motion of the menisci, and as long as the menisci move within a single tube this is straightforward. However, there will be menisci reaching the ends of the tubes. This is dealt with as follows for every node separately. All the menisci which have proceeded past the edges of the tubes neighboring a node, and so have entered the node, represent incoming volume of one of the fluids. In order to have volume conservation the same amount of volume must leave the node. This is done by adding new menisci in the neighboring tubes in which the flow is away from the node. The positions of these new menisci are such that volume is preserved. This is the essential part of these rules of motion. However, the rules which are actually used are a little more complicated, mainly due to computational technicalities. They have been presented in detail before\cite{KAH00}, and they are not necessary for the understanding of results in this paper.

Many authors have studied invasion under a constant applied pressure. In the  work by Aker \emph{et al.} constant flux invasion has been considered. By solving the flow field for two different globally applied pressures giving two different fluxes, one can calculate the pressure that would give the desired flux. We do not go into details here since this is described in detail before\cite{AMHB98}. We will just remark for now that all simulations presented in this paper are done with constant flux rate.

Network models based on Washburn's equation was introduced by the Payatakes group, see e.g.\cite{CP91} and references therein. Their work related to steady-state two-phase flow is of particular interest\cite{CP96,VP01}. A detailed comparison of the two lines of modeling is not our aim, but we give some general remarks. Our model is a little more coarse in two ways. First, the resolution of pore space and of the fluid in the pore space is lower in our work since it is a tube model, and since only two menisci are allowed in each tube. Second, film flow is not included, neither explicit as network components or via the use of coefficients for, e.g., coalescence of oil blobs. Also, we have chosen to work with fixed values for a larger number of parameters. In other words, we present generic rather than specific use of the model in this piece of work. Finally, we emphasize that our model use biperiodic boundary conditions to reach steady-state flow. A short description follows.

\subsection{Boundary conditions}
\label{sec:boundaryconditions}
Invasion simulations go on until the invading fluid reaches the outlet. If they were to go on further, they change character since one fluid is percolating the system. We wish to address the question of what happens in a system far from inlets and outlets, i.e. given a very large system, we take out a small piece somewhere in the middle and study its properties. 

This is done by adjoining the inlet row and outlet row so that the fluid that flows out of the last row enters the first row. In practice this works in such a way that the simulation can go on for ever, regardless whether one fluid percolates the system or not. In a sense having biperiodic boundary conditions make the system infinite. However, the system is closed, and there is a fixed volume of each of the fluids in the system. Thus each simulation takes place at constant saturation equal to the one of the initial configuration. We return to this point when presenting results from simulations in section \ref{sec:simulations}.

These boundary conditions can be visualized by considering the system to be the surface of a torus. The flow is driven around the torus by a numerical trick. This method has been used in connection with random resistor networks\cite{Roux}. Briefly, one can say that there is an invisible line dividing the system. Looking over this line, there is a pressure fall or jump, which is the driving force of the flow. All the tedious details are given in the previous piece of work\cite{KAH00}, none of which are necessary to know in order to understand the contents of the present work.

\section{Simulations}
\label{sec:simulations}

The essence of the model porous network is that it is situated on the surface of a torus. In other words the applied boundary conditions are biperiodic, which makes the system closed. By means of the pressure fall technique, the overall flow is around the torus, while the total volumes of the two fluids remain constant within the system.

\begin{figure}
\begin{tabular}{c}
\includegraphics[width=7cm]{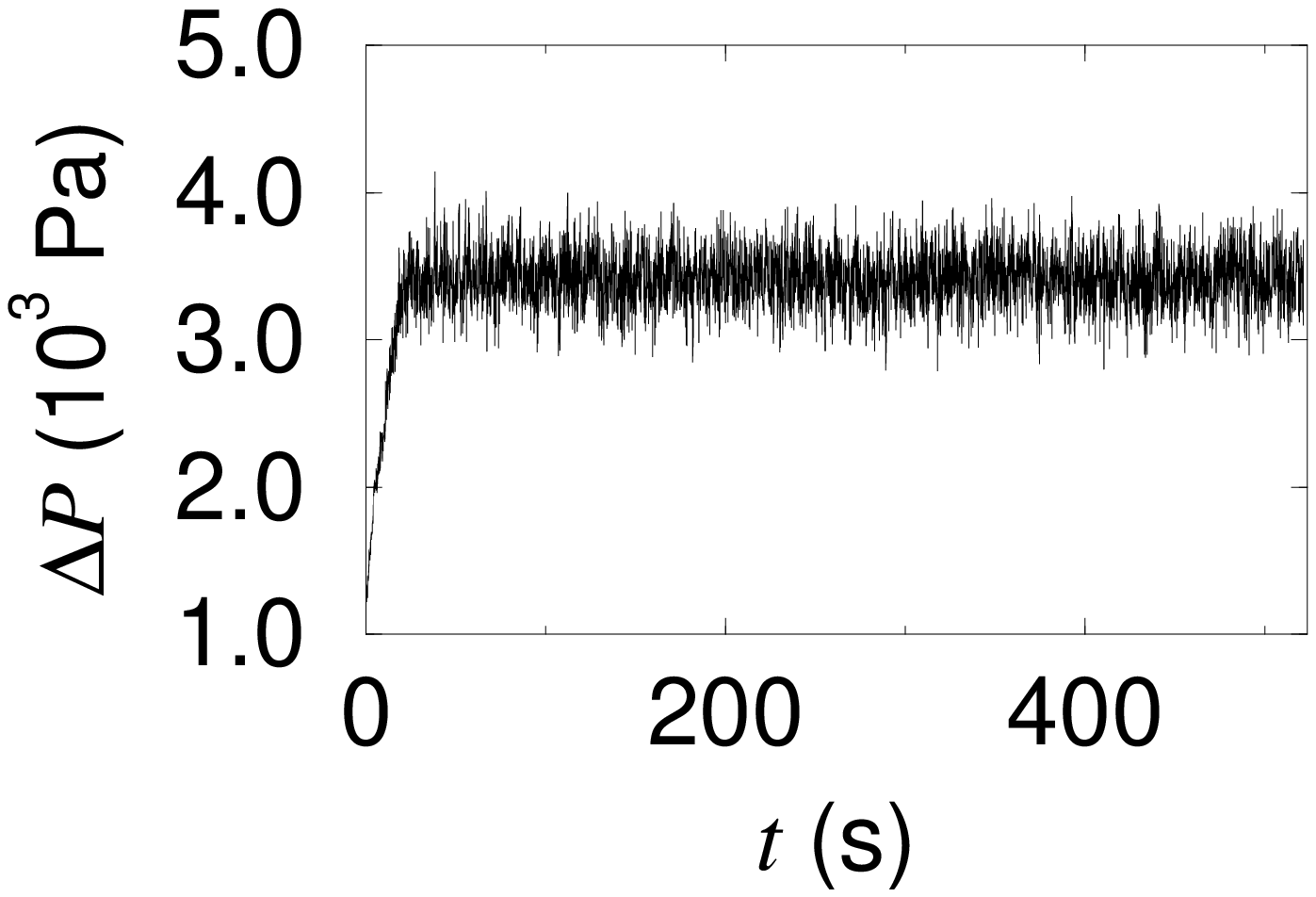} \\
(a) Global pressure \\ \\
\includegraphics[width=7cm]{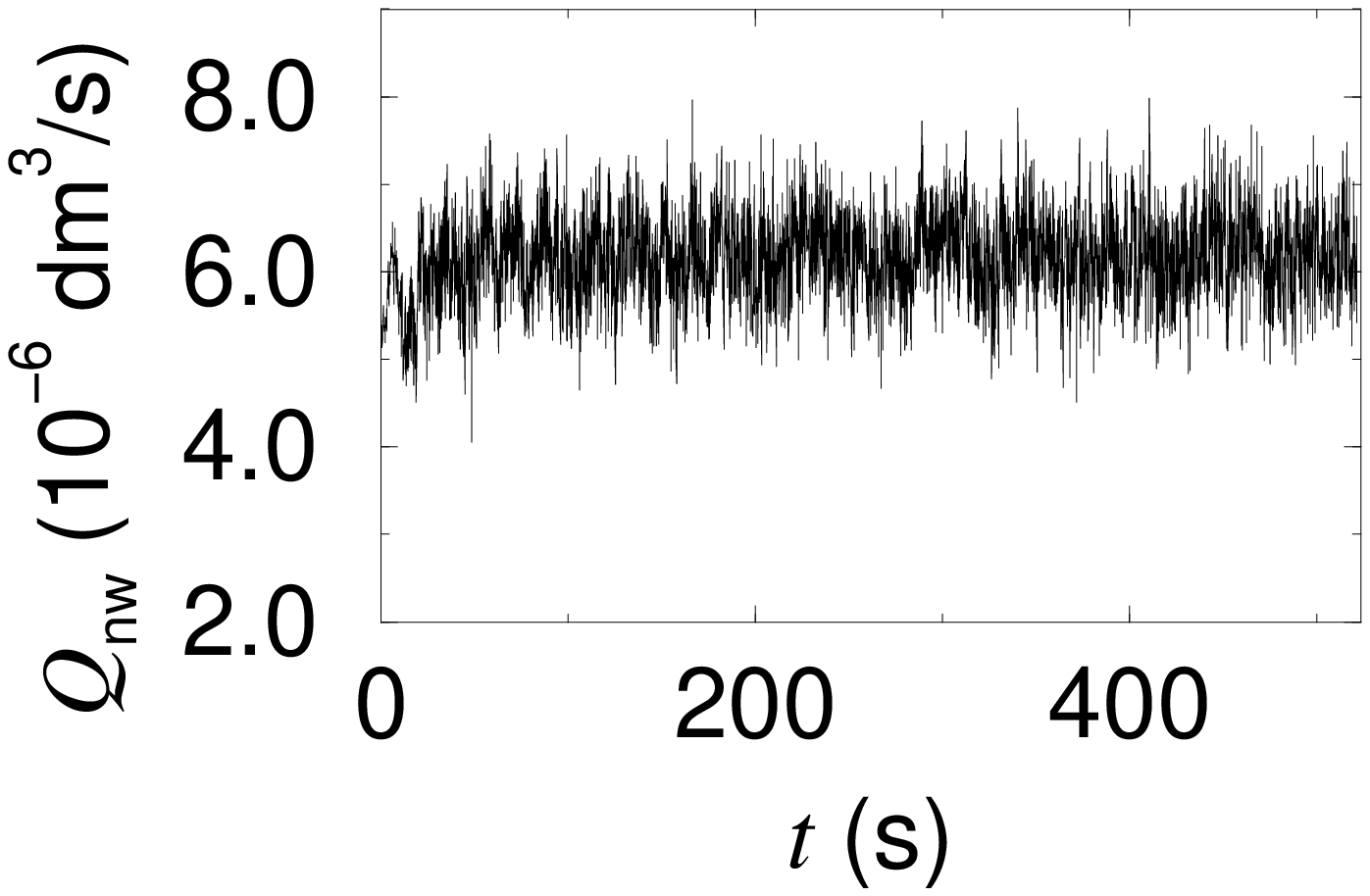} \\
(b) Nonwetting flux
\end{tabular}
\caption{The figures show the typical time evolution of (a) global pressure, and (b) flux of nonwetting fluid, in a system of size $20\times 40$ nodes. The nonwetting saturation is approximately $35\%$, the capillary number is $C_{\rm a}=3.2\times 10^{-3}$, and the total flux through the system is $Q_{\rm tot}=14.0\times 10^{-6}{\rm dm}^3/{\rm s}$. On the average the fluid travels almost 17 times around the system during this simulation.}
\label{fig:prestime}
\end{figure}

Equally important is the fact that the simulations, which are presented here, are done by keeping the total flux around the torus constant. As time evolves the fluid distribution changes and local capillary pressures change. In order to keep the total flux constant, the globally applied pressure fall must change in every time step. A sample of global pressure versus time is given in Fig. \ref{fig:prestime}(a).

We wish to draw attention to two aspects of the nature of this curve. The first is that it appears noisy. On a short time scale local redistributions of the fluids and changes in the capillary pressures lead to a change in the overall pressure which must be applied in order to produce a given flux. Also one should remember that the time axis is very compressed. The fluctuations in pressure are not achieved in one single time step, but rather in the order of one hundred time steps.

The second aspect is that on a longer time scale the systems have a transient part and a steady part. The transient part depends on the initial conditions, while in general we have found that the steady part is independent of the initial conditions. After the systems reach the steady state, the properties of this state are typically the time averages over the variables: global pressure, the flux of each of the fluids, the number of interfaces, velocity distributions, and others.

Fig. \ref{fig:prestime}(b) shows the nonwetting flux $Q_{\rm nw}$ as a function of time. The data in \ref{fig:prestime}(a) and \ref{fig:prestime}(b) are from the same simulation. We see how the nonwetting flux also has a transient and a steady part. More convenient is the nonwetting fractional flow, which is defined as the flux of the nonwetting fluid through the system divided by the total flux: $F_{\rm nw}=Q_{\rm nw}/Q_{\rm tot}$. To some extent the fractional flow property has been studied before. Previous results will be summarized shortly.

\subsection{Fractional flow and pressure curves}
\label{sec:fraccurves}

There are many possible parameters of the model system. We will generally use the saturation as the independent variable in our plots. The nonwetting saturation is defined as the volume of nonwetting fluid divided by the total volume, $S_{\rm nw}=V_{\rm nw}/V_{\rm tot}$. The other possible parameters are the viscosities of the fluids, the total flux, the inter-facial tension, and system size.

In this work we have restricted ourselves to the case where the two fluids have equal viscosity. The viscosity and some other parameters are combined in the capillary number
\begin{equation}
\label{eq:capnum}
C_{\rm a} = \frac{\mu Q_{\rm tot}}{\gamma \Sigma},
\end{equation}
where $\mu$ is the viscosity, $Q_{\rm tot}$ is the total flux, $\gamma$ is the inter-facial tension, and $\Sigma$ is the cross-sectional area of the system. Physically this number is the ratio between typical viscous forces and capillary forces within the system. We have used a fixed value for the inter-facial tension, $\gamma=30.0 {\rm mN/m}$. The cross-sectional area is approximately $\Sigma=0.145 {\rm cm}^2$ for the system size which is used. We have further chosen the value of the viscosity to be $\mu=0.1 {\rm Pa\ s}$. Actually the total flux is the only one of these parameters which we vary. This is no limitation as we will discuss shortly.

\begin{figure}
\begin{tabular}{c}
\includegraphics[width=7cm]{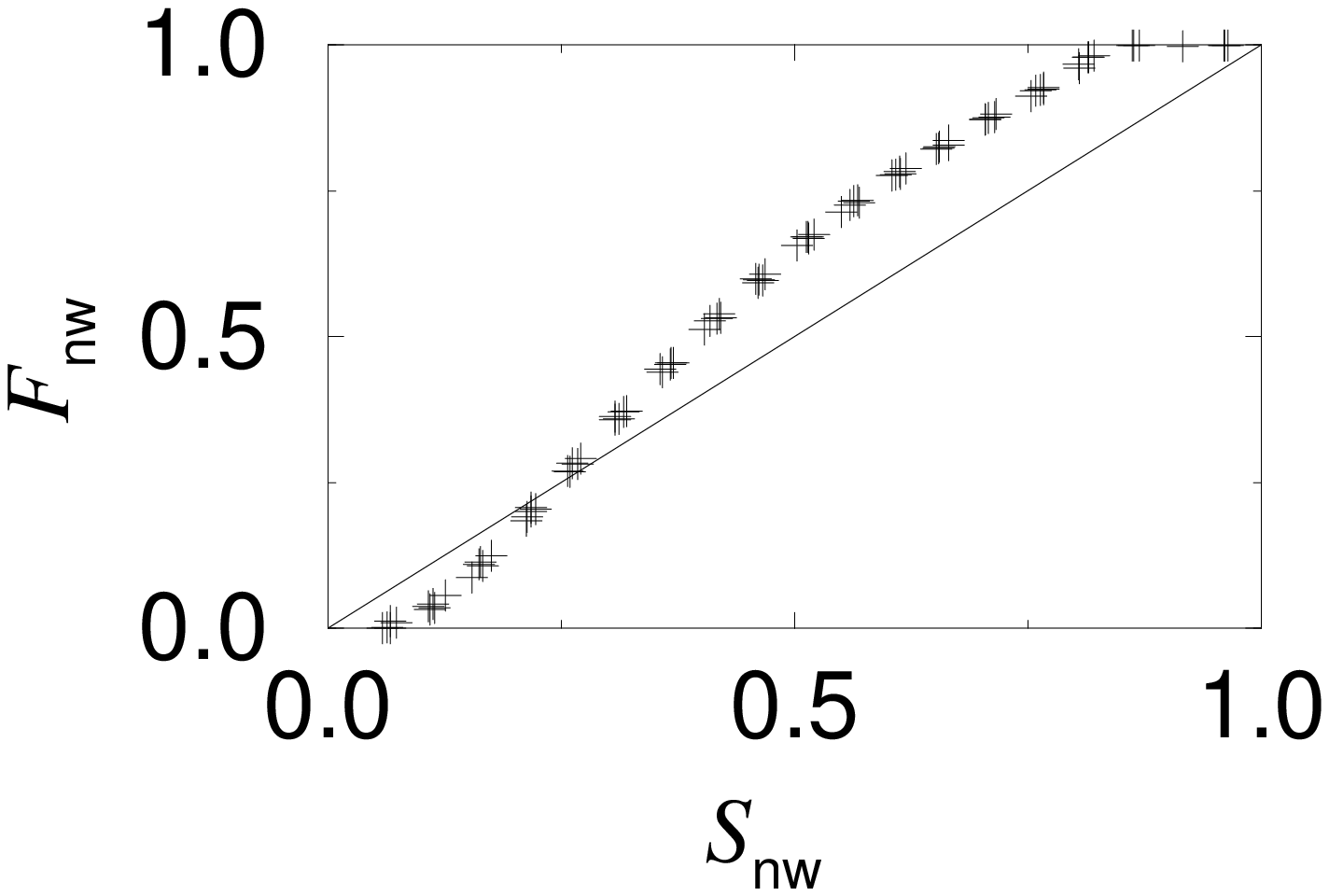} \\
(a) Nonwetting fractional flow \\ \\
\includegraphics[width=7cm]{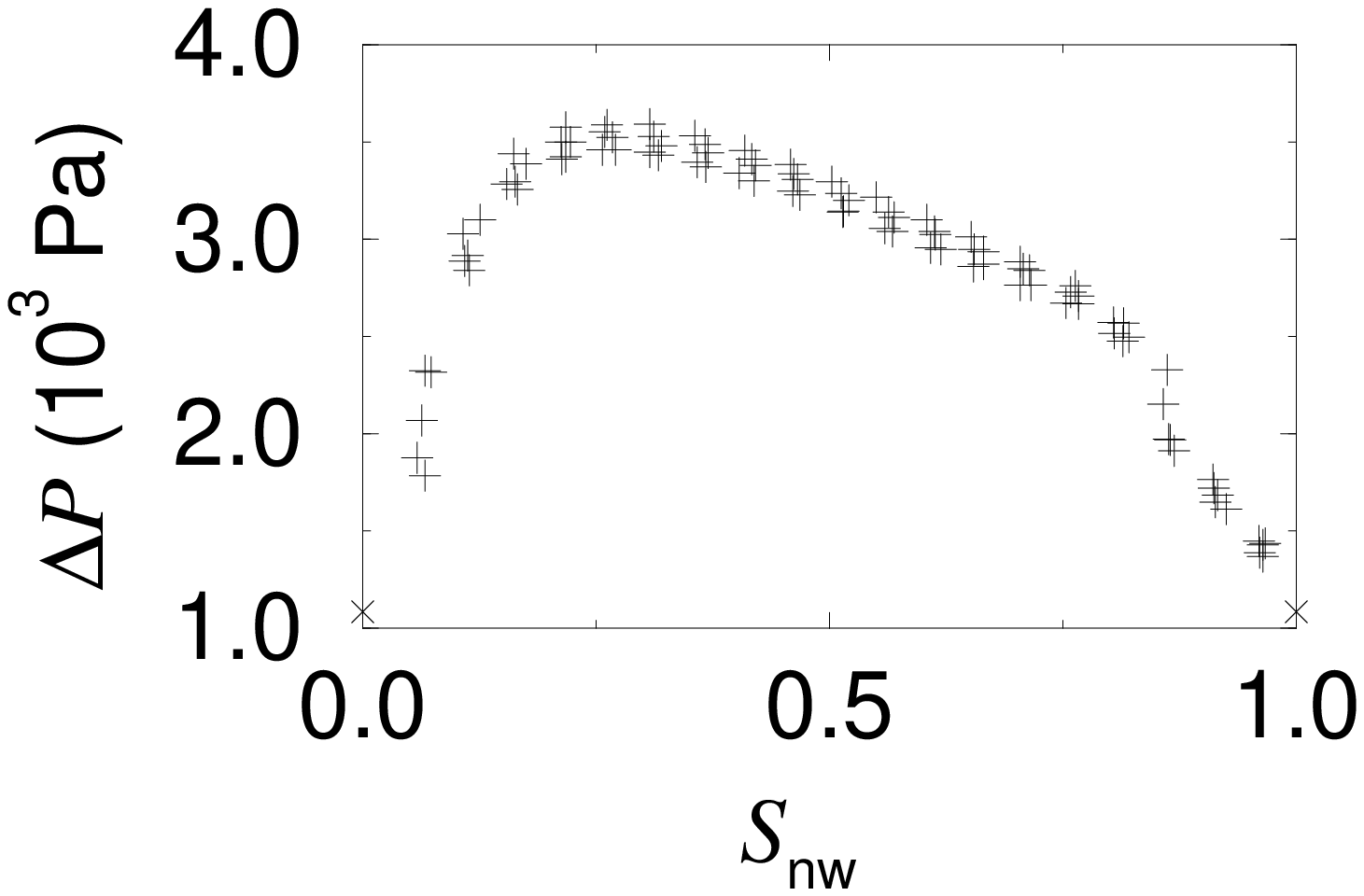} \\
(b) Total pressure
\end{tabular}
\caption{The plots show average values of (a) fractional nonwetting flow and (b) global pressure, where the averages are taken in the steady part. The capillary number is $C_{\rm a}=3.2\times 10^{-3}$. We see clearly how these values depend on the nonwetting saturation. The simulations are run on five different geometries generated by different random seeds.}
\label{fig:fracpres}
\end{figure}

For the capillary number, $C_{\rm a}=3.2\times 10^{-3}$ the nonwetting fractional flow as a function of nonwetting saturation is given in Fig. \ref{fig:fracpres}(a). The corresponding global pressure is shown in Fig. \ref{fig:fracpres}(b). Here it should be noted again that the values of the fractional flow and global pressure are the time averages in the steady state.

The very important question of how these results depend on the system size was examined before\cite{KAH00}. When all other parameters were held constant, except for system size and saturation, the fractional flow as a function of saturation showed to be independent of the system size. The examined system sizes were $20\times 40$ nodes, $20\times 80$, and $40\times 80$.

The interpretation of the size independence is that the system is large enough to be in the asymptotic limit. Of course, this might happen to be untrue for some possible parameter sets of the system, but for the sets used in the present work it is true. All simulations which are presented here are therefore done with system size $20\times 40$ nodes.

The system size dependency of the global pressure is slightly more complicated. The pressure is applied in such a way that there is a pressure gradient along one of the system axes. Any change in size sideways, that is perpendicular to the gradient, does not alter the pressure curves. However, increasing the length along the pressure gradient with a factor results in a global pressure which is increased by the same factor. This is just another way of saying that it is the gradient in global pressure, which must be applied to give a desired total flux, that is independent of system size. In this work we consider only one system size so we prefer to use $\Delta P$ for the global pressure drop.

The points in the plots come from five different geometries. They are drawn from the same distributions with different random seeds. The specific permeability of the porous medium, denoted by $k$, is related to the applied pressure and the total flux for a single phase, Darcy's law,
\begin{equation}
  \frac{Q}{\Sigma} = \frac{k}{\mu} \frac{\Delta P_{\rm s}}{L},
\label{eq:darcy}
\end{equation}
where $L$ is the length of the system along the pressure gradient, and the subscript on $\Delta P_{\rm s}$ denotes single phase pressure. This is a geometrical property and will therefore vary slightly for the five geometries. In fact this means that the pressure curves in Fig. \ref{fig:fracpres}(b) should be five slightly different curves. The variations in the specific permeability are of the order of a couple of percents. The five curves are scaled according to this difference. The average position of the curves is kept constant under the scaling. We note that this implies that the single phase points, 0\% and 100\% saturation, are shown as single points. The rest of the curves are not perfectly overlapping due to the statistical variations.

Subsequently we will take this scaling one step further. For illustrative purposes the physical dimension of pressure was used in Fig. \ref{fig:fracpres}(b). However, it will be more convenient to work with a normalized pressure. This is done by giving the pressure in units of the single phase pressure, i.e., letting the axis be $\Delta P/\Delta P_{\rm s}$. This is straighforward, but note that the single phase pressure is a function of the capillary number. It follows from Eqs. (\ref{eq:capnum}) and (\ref{eq:darcy}) that $\Delta P_{\rm s} \propto C_{\rm a}$.

The fact that the capillary number, as it is defined in Eq.(\ref{eq:capnum}), serves as the relevant combined parameter was also established in the previous piece of work\cite{KAH00}. This means that one can make changes to all the variables in Eq. (\ref{eq:capnum}), but as long as the capillary number is preserved the shape of the fractional flow curve will be the same.

The characteristics of the fractional flow curve are as follows. Below a certain nonwetting saturation, approximately $5\%$ for $C_{\rm a}=3.2\times 10^{-3}$, we observe that the nonwetting fractional flow is essentially zero, meaning that only the wetting phase flows. Conversely above a certain nonwetting saturation, $85\%$ for $C_{\rm a}=3.2\times 10^{-3}$, only the nonwetting phase flows. Between these values both phases flow.

Imagine there had been no capillary forces between the two phases. Then both phases would have flown equally easy through the porous network. The nonwetting fractional flow would then be exactly equal to the nonwetting saturation. This is illustrated by the diagonal line in Fig. \ref{fig:fracpres}(a). The effect of having interfaces with capillary pressures can be thought of as the change from the ideal diagonal to the obtained curve. For low nonwetting saturations the fractional nonwetting flow is lower than the saturation, i.e. under the diagonal. For high nonwetting saturation the nonwetting fractional flow is higher than the saturation, i.e. above the diagonal. Roughly we can say that the phase of which there is more volume, gains and flows more than one naively might expect. At some point the curve cross the diagonal, and this is the point where neither phase gains compared to its volume fraction. This point is not at $50\%$ saturation, and it is clear that there is an asymmetry between the wetting and nonwetting phases.

The pressure curve in Fig. \ref{fig:fracpres}(b) has two points plotted with '$\times$', at $0\%$ and $100\%$ saturation. These are single phase pressures, and since the two fluids have the same viscosity the pressures are equal. For all other saturations, two phases are present within the system. There are interfaces between the phases with capillary pressures. Motion of two fluids with interfaces through tubes requires more global pressure than for the case of a single fluid. One might imagine that in some cases some or all of the present interfaces do not move, which is typically the situation when only one fluid flows. Still the global pressure must be higher than for a single fluid because the nonmoving interfaces block out tubes, reduce possible pathways, and hence reduce the specific permeability of the medium. As we can see from the figure, the global pressure is higher in the entire two-phase region than in the single-phase points.

The general nature of the pressure curve is that with increasing saturation the pressure increases monotonically to a maximum value. Thereafter it decreases monotonically. The curve in Fig. \ref{fig:fracpres}(b) is generally smooth except at nonwetting saturation of approximately $85\%$ where the pressure drops rather rapidly. We observe for now that this point is coinciding with the point were the nonwetting fractional flow becomes unity. Also the character of the curve changes at this point. To the left it is curved while to the right it is nearly a straight line.

\subsection{The dependency on $C_{\rm a}$}
\label{sec:cadep}

\begin{figure}
\begin{tabular}{cc}
\includegraphics[width=4cm]{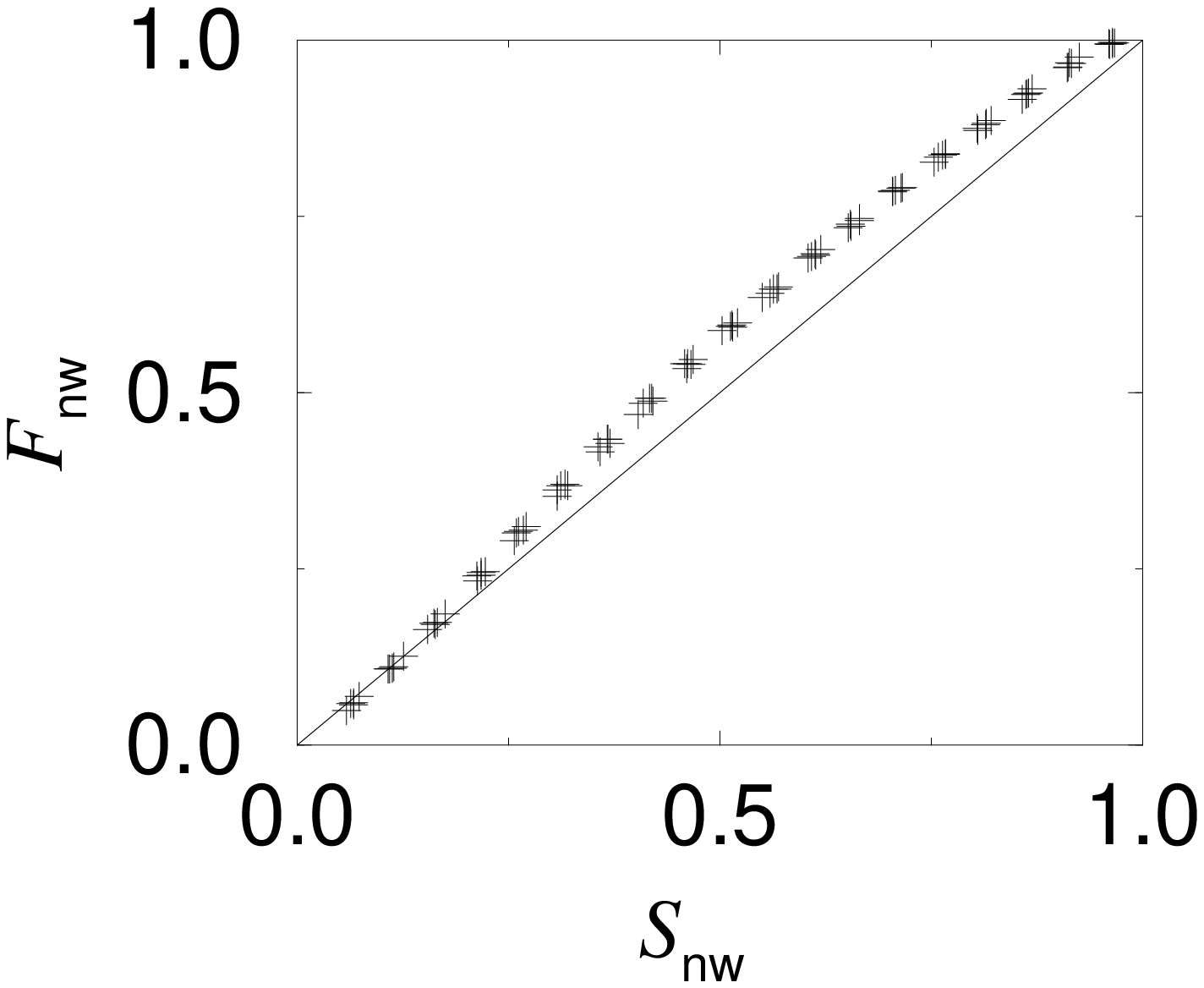} &
\includegraphics[width=4cm]{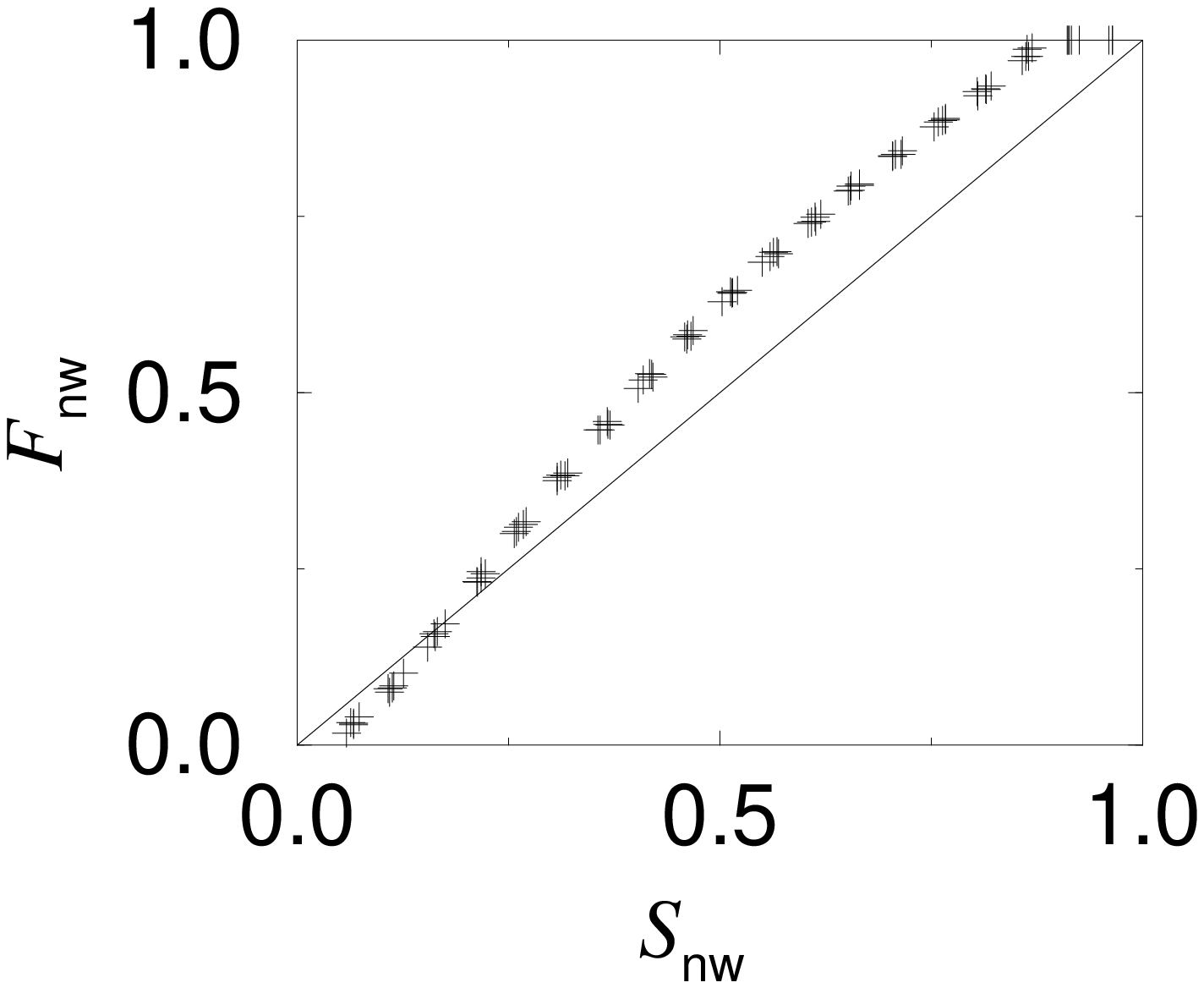} \\
(a) $C_{\rm a} = 3.2\times 10^{-2}$ &
(b) $C_{\rm a} = 1.0\times 10^{-2}$ \\
 & \\
\includegraphics[width=4cm]{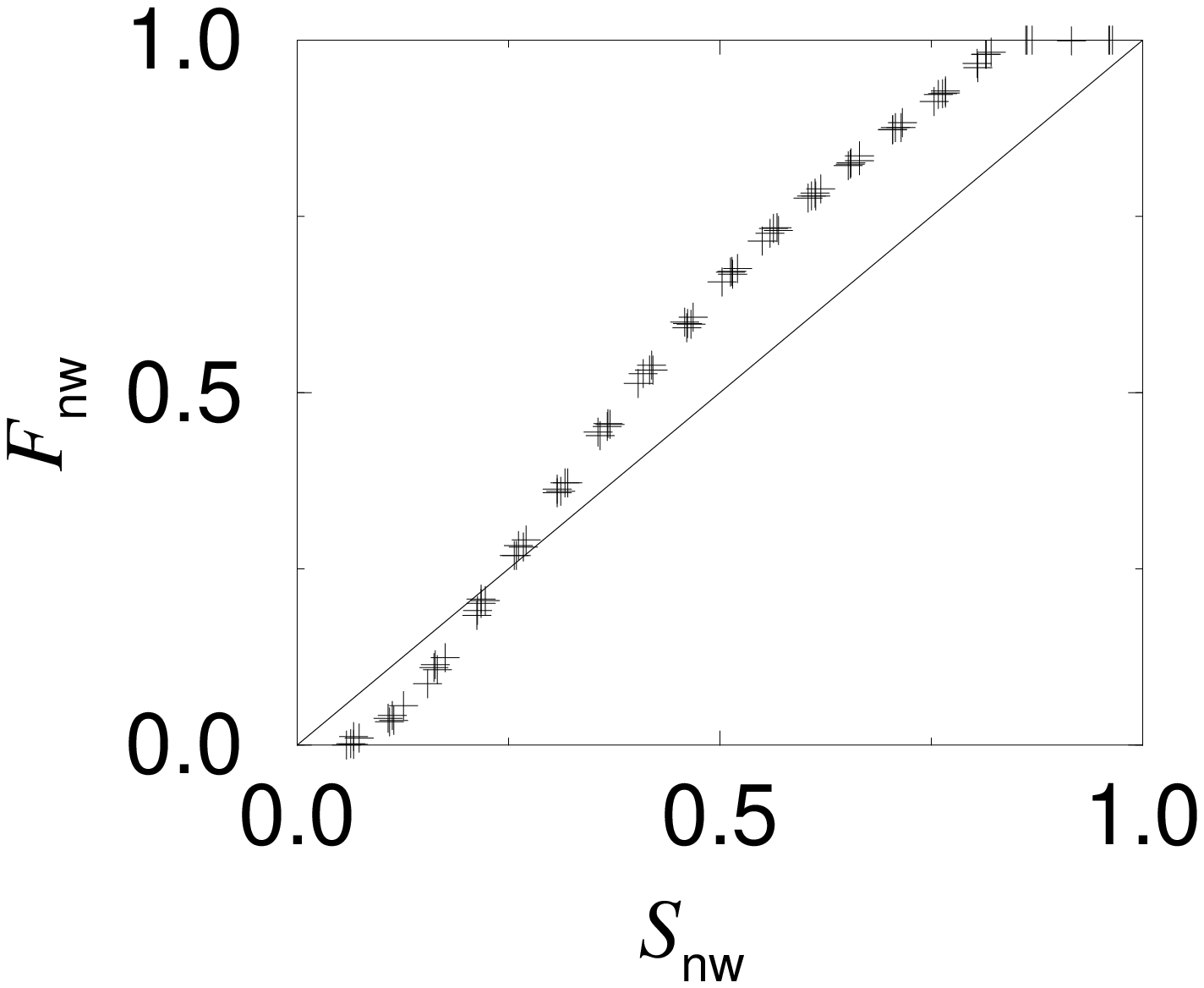} &
\includegraphics[width=4cm]{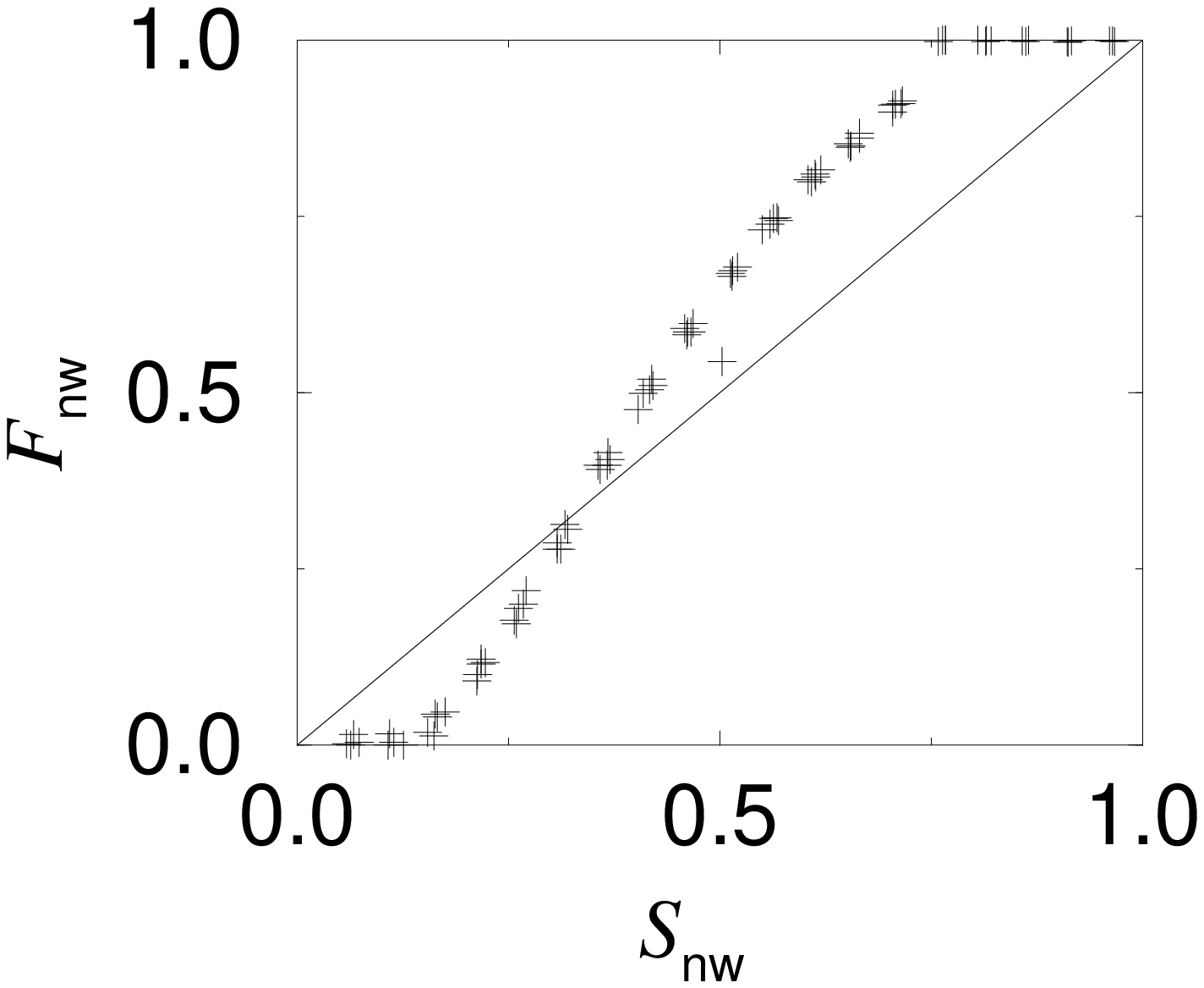} \\
(c) $C_{\rm a} = 3.2\times 10^{-3}$ &
(d) $C_{\rm a} = 1.0\times 10^{-3}$ \\
 & \\
\includegraphics[width=4cm]{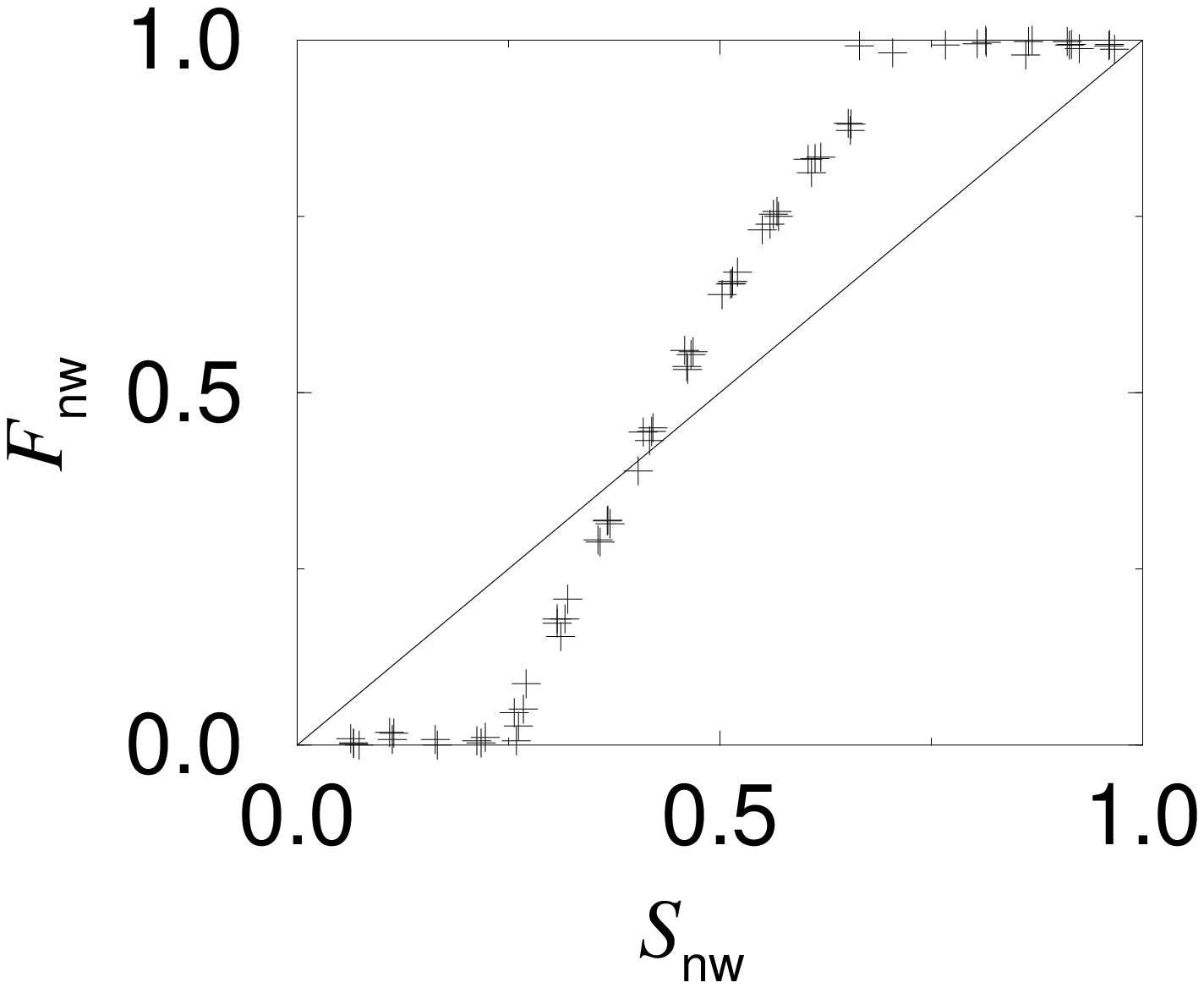} &
\includegraphics[width=4cm]{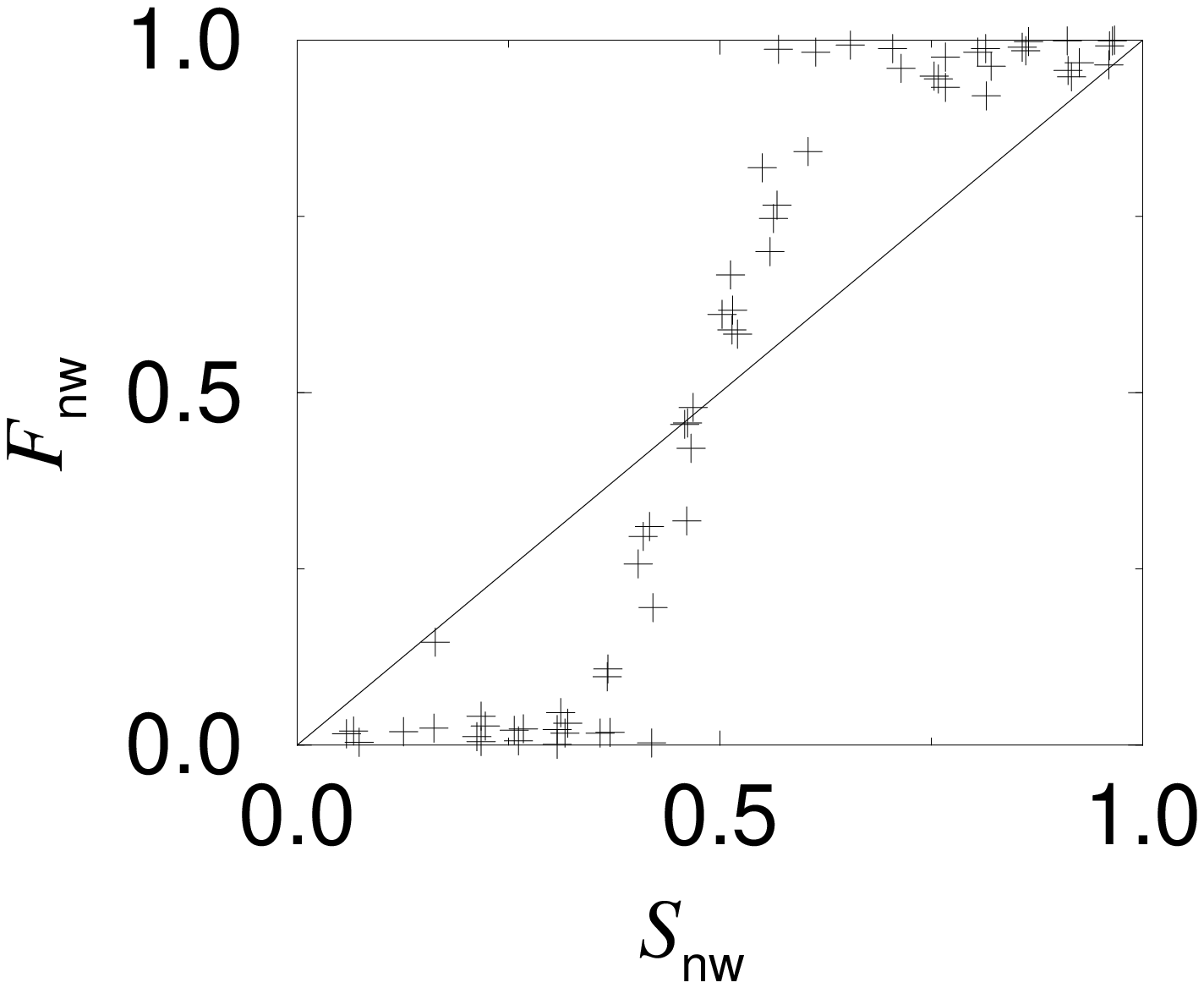} \\
(e) $C_{\rm a} = 3.2\times 10^{-4}$ &
(f) $C_{\rm a} = 1.0\times 10^{-4}$
\end{tabular}
\caption{The figures show the nonwetting fractional flow as function of nonwetting saturation for six different capillary numbers. For high capillary numbers the curve is close to the diagonal, while for low capillary numbers the curve is S-shaped. The range of saturations for which both phases are mobilized decreases with decreasing capillary number.}
\label{fig:fflows}
\end{figure}

\begin{figure}
\begin{tabular}{cc}
\includegraphics[width=4cm]{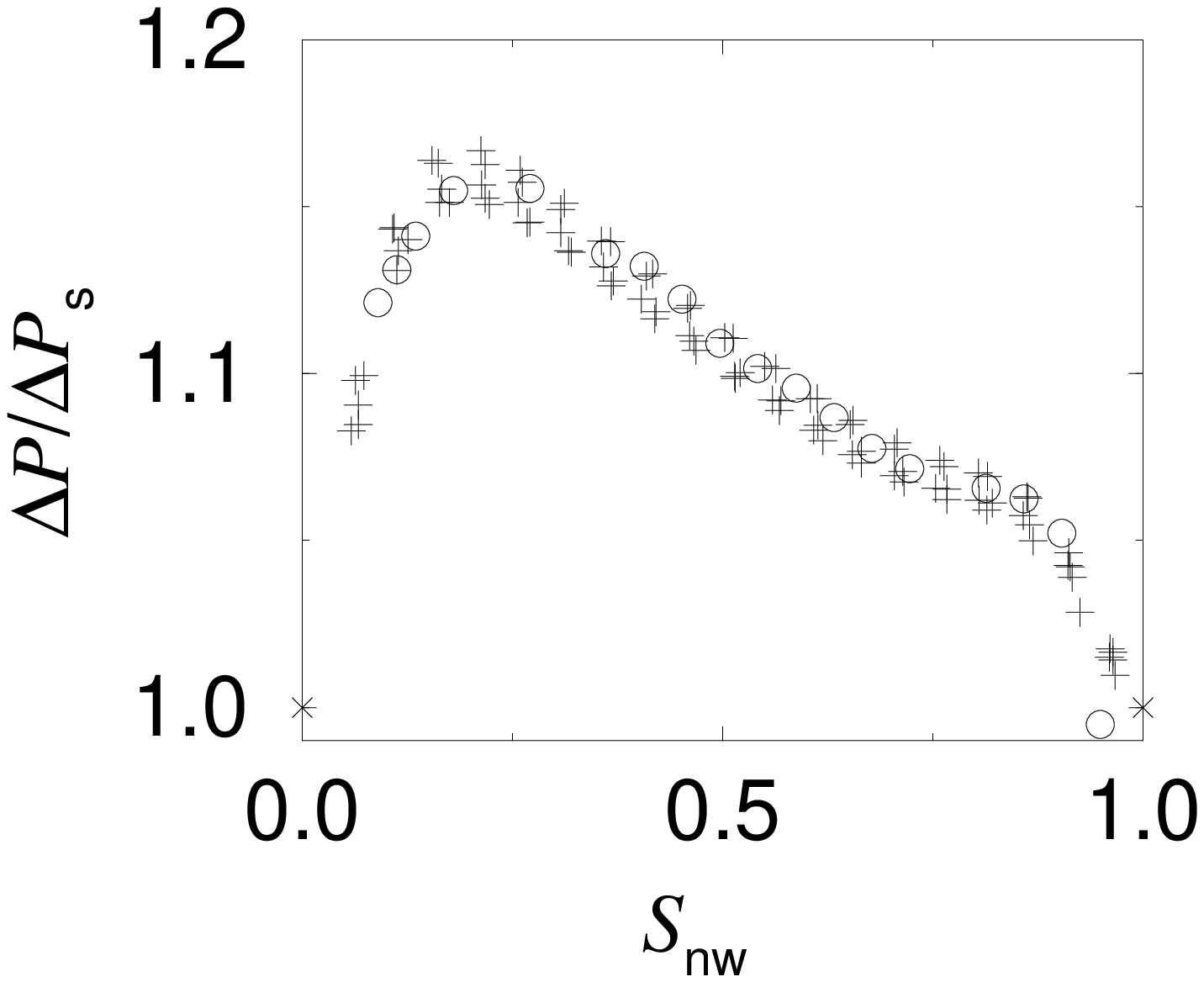} &
\includegraphics[width=4cm]{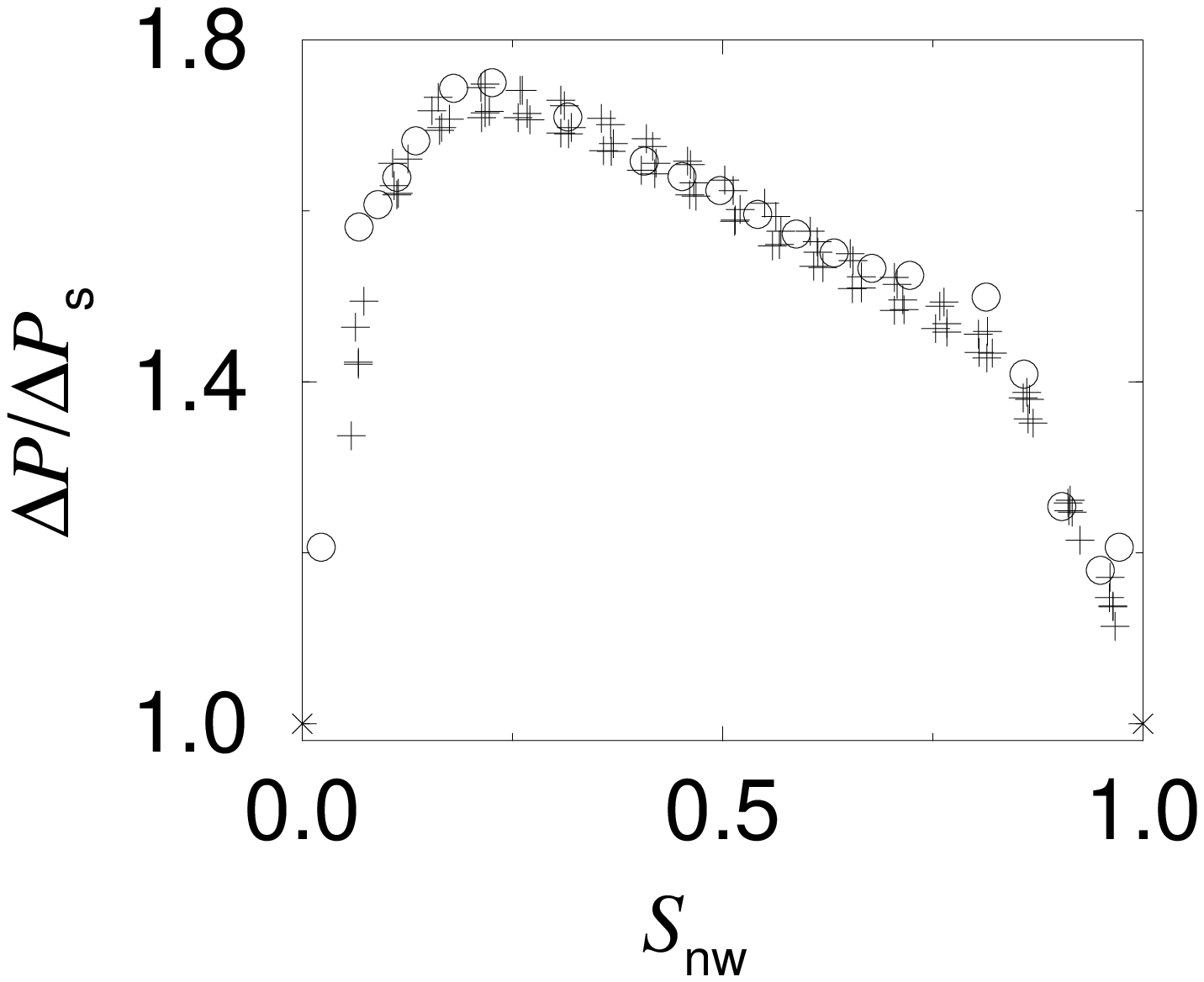} \\
(a) $C_{\rm a} = 3.2\times 10^{-2}$ &
(b) $C_{\rm a} = 1.0\times 10^{-2}$ \\
 & \\
\includegraphics[width=4cm]{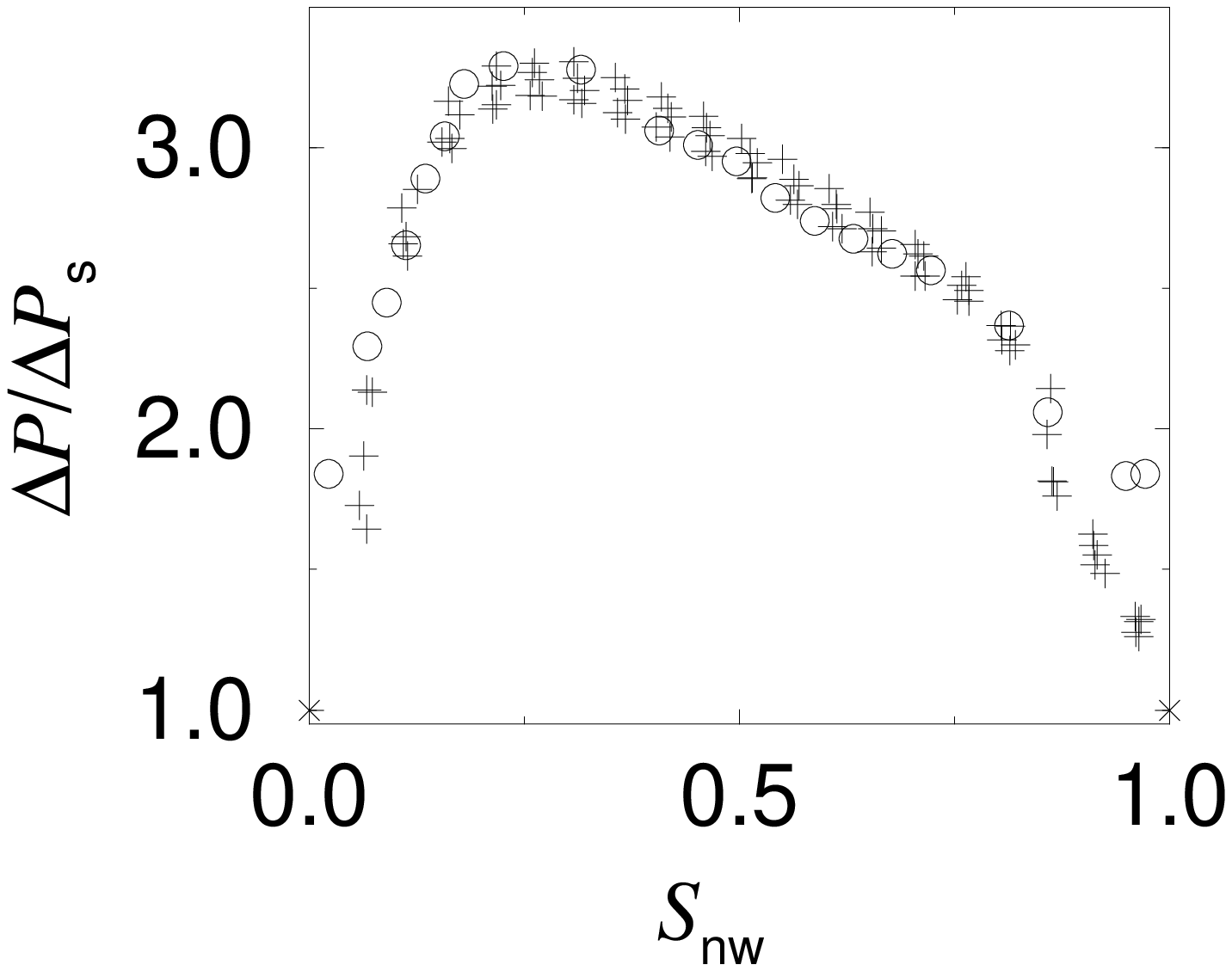} &
\includegraphics[width=4cm]{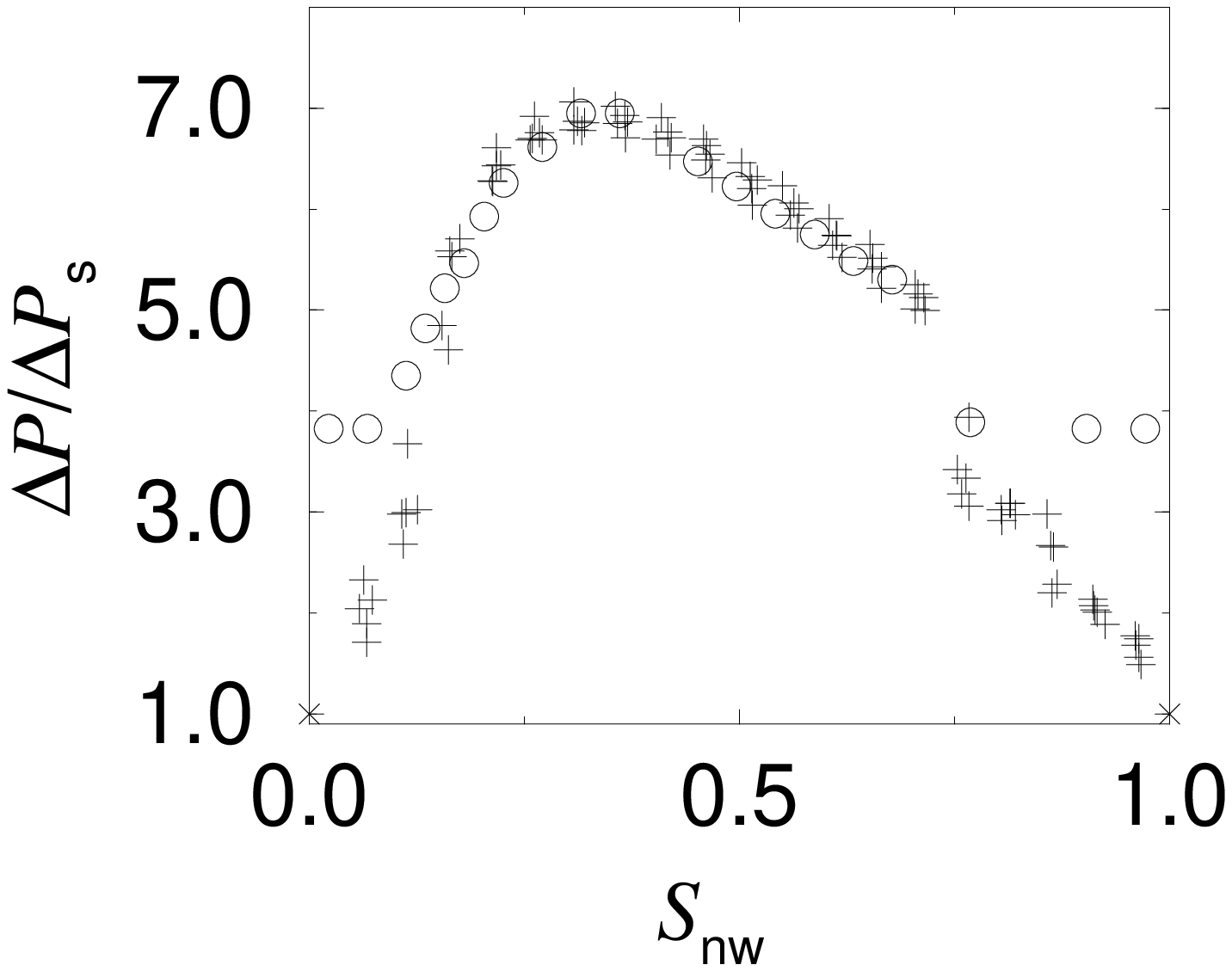} \\
(c) $C_{\rm a} = 3.2\times 10^{-3}$ &
(d) $C_{\rm a} = 1.0\times 10^{-3}$ \\
 & \\
\includegraphics[width=4cm]{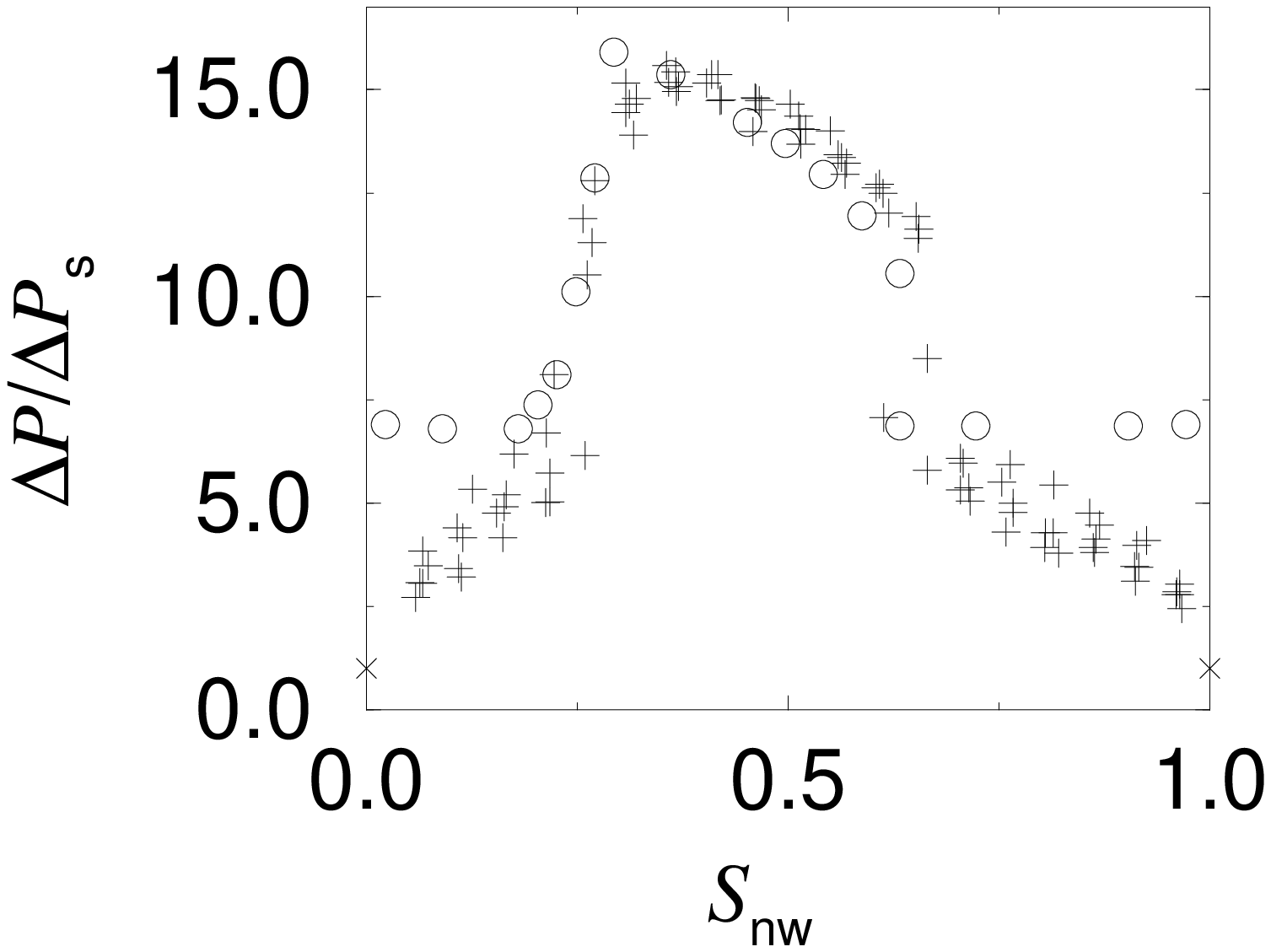} &
\includegraphics[width=4cm]{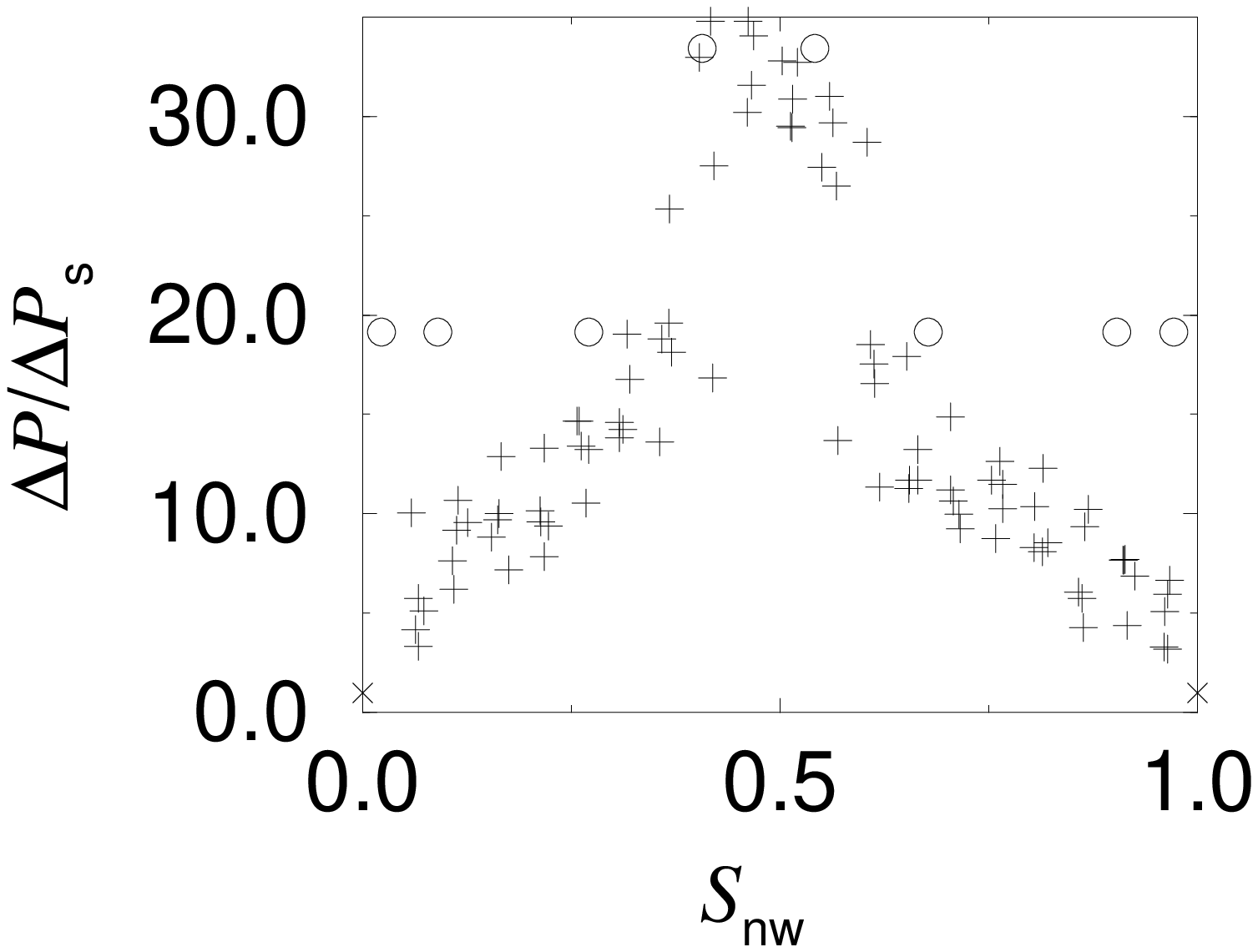} \\
(e) $C_{\rm a} = 3.2\times 10^{-4}$ &
(f) $C_{\rm a} = 1.0\times 10^{-4}$
\end{tabular}
\caption{The figures show the global pressure, '$+$', as a function of nonwetting saturation for six different capillary numbers. The global pressures are normalized with respect to the single phase pressure $\Delta P_{\rm s}$ as defined in (\ref{eq:darcy}). Recall that for each capillary number the total flux is fixed to a constant value. The curves correspond to the fractional flow curves in Fig. \ref{fig:fflows}, and the derivate of the fractional flow curves are included here in a scaled version, '$\circ$'. This scaling and the relationship between the curves are treated in subsection \ref{sec:presanddfds}.}
\label{fig:globpres}
\end{figure}

Now we have learned that the capillary number serves as the relevant parameter for the fractional flow versus saturation curves. For two fluids having equal viscosity we have performed simulations for six different capillary numbers, which is shown in Figs. \ref{fig:fflows} and \ref{fig:globpres}. The system size is $20 \times 40$ in all simulations. This size is sufficiently large to be in the asymptotic limit, but still so small that simulations can be done within reasonable amounts of time. For each capillary number we have chosen 19 different saturations from approximately $5\%$ up to $95\%$ in steps of $5\%$. Further we have used five random seeds which provide five different geometries.

The shape of the fractional flow curve depends strongly on the capillary number. For high capillary numbers the curve is approaching a straight line. Intermediate capillary numbers, as for the sample which was discussed in the previous subsection, the curve has a clear S-shape. For lower capillary numbers the curve seems to approach a step function.

The curves can be divided into three regions. The first being for low nonwetting saturation where the nonwetting fractional flow is zero. For the two highest capillary numbers this region is almost vanishing. As the capillary number is lowered, this region grows in size. The third region is for high nonwetting saturations where the nonwetting fractional flow becomes unity. The same general behavior is valid here as for the first region. It becomes wider with decreasing capillary number. The difference is that for a fixed capillary number the region with unity nonwetting fractional flow is larger than the region with zero nonwetting fractional flow. This means that there is not perfect symmetry between the two phases.

The second region is the central part where both phases are mobilized. The width of this region decreases with decreasing capillary number. In the limit of infinite capillary number, negligible capillary forces, the curve will approach the diagonal and thus span all of the saturation range. In the opposite limit of small capillary number, it is reasonable to expect that the curve will approach a step function, although we have not explicitly checked this for lower capillary numbers than presented here. Between these limits there is a range of capillary numbers where the central parts of the curves have interesting structure.

As it was pointed out in the previous subsection for $C_{\rm a}=3.2\times 10^{-3}$, the curves lie above the diagonal for large nonwetting saturations and below for small nonwetting saturations. The interpretation is that the system favors transport of the phase of which there is more volume. This is not exactly true, since the crossover from favoring one phase to the other is not at $50\%$ nonwetting saturation, but at a somewhat smaller value. Again the system is not perfectly symmetric in the two phases. Actually the crossover point is a function of the capillary number. For very high $C_{\rm a}$ the crossover point is close to zero. It approaches $50\%$ when the capillary number decreases. This situation corresponds to the problem of bond percolation in two dimensions, for which the percolation threshold is known to be $s_{\rm c}=1/2$\cite{SA92}. In percolation theory $s_{\rm c}$ is a critical bond probability, and it corresponds to a critical saturation in our problem. For small systems, history may evolve in such a way that the systems have a saturation other than $50\%$ even though only one phase flows. However, on the average, or in an infinite system which is self-averaging, the limiting value is $50\%$. We have listed the values of the crossover point in Table \ref{thetablei}.

Corresponding to the fractional flow curves in Fig. \ref{fig:fflows} are the global pressure curves in Fig. \ref{fig:globpres}. Also here the curves depend strongly on the capillary number. It was pointed out in the previous subsection for $C_{\rm a}=3.2\times 10^{-3}$ that the curve increases monotonically to a maximum value and thereafter decreases monotonically. This is true for the entire range of capillary number which we have examined. The position of the maximum value increases with decreasing capillary number. The values are listed in Table \ref{thetablei}.

The pressure curves can be divided into three regions just like the fractional flow curves. At least for the three lowest capillary numbers, Fig. \ref{fig:globpres}(d)-(f), their boundaries are clear and distinguishable. The first region is the one were the pressure increases linearly with nonwetting saturation from the value at zero nonwetting saturation. For the three highest capillary numbers, Fig. \ref{fig:globpres}(a)-(c), this region is almost vanishing.

The third region is where the curves decrease linearly with increasing nonwetting saturation towards the value of pressure at unity nonwetting saturation. This region can be seen in the entire range of capillary numbers. By inspection we see that the region boundaries are the same in the pressure curves as for the fractional flow curves. Thus the same comments regarding the width of the regions are valid for the pressure curves.

The second region is the central part of the curve. It is curved and has nontrivial structure. For the lower capillary numbers the pressure increases abruptly at the boundaries of the region. In this region both phases are mobilized. The immediate conclusion is that more pressure is needed when both phases are mobilized, as interfaces then are mobilized. The relation between the fractional flow and the global pressure in this region will be the subject of the next subsection.

The first and the third region share several properties. At their outer boundaries, $0\%$ and $100\%$ respectively, the values of the global pressure are equal, and equal to the single phase pressure. Also at the inner boundaries of these regions the global pressures have the same value. That is to say, as far as both inner boundaries are distinguishable, we can make this observation. Physically the inner boundaries are just at the saturation values where there is an onset of mobilization of the other phase. The fact that the pressures are the same here is the same as saying that in this respect there is symmetry between the two phases.

Within these regions one of the phases is immobilized. Still there is a linear dependency of the global pressure on the saturation within the regions. The reason for this change in pressure cannot come from the increased pressure necessary to move an increased number of interfaces since the interfaces generally are pinned to fixed positions when only one phase flows. When there is an increased volume of the phase which is immobilized, then there is less available volume for the other phase to move in. This is a geometrical constraint. The specific permeability of the single phase which flows will decrease within these two regions as the saturation of other phase increases. Thus more and more pressure is needed to maintain a constant total flux.

This qualitative explanation does not account for the fact that these parts of the curves are straight lines. If the immobilized phase were pinned to positions in the network which were more or less random, then to a first approximation a linear increase in the saturation of the immobile phase would lead to a linear decrease in the effective system size for the mobile phase. In turn that would lead to a linear increase in global pressure.

The next question is why the third region is wider than the first region. According to the first approximation reasoning above, they should be equal. Going a little beyond this approximation, we know that a bubble of nonwetting fluid is more likely to get pinned around a node. Its most stable position is when it is bounded by menisci in all neighboring tubes, and when they are in the half parts of the tubes which are closer to the node. Whatever direction the bubble is pushed it will have to pass the threshold in the respective tube. Bubbles of wetting fluid have opposite preference. Their most stable position is within a tube. With two bounding menisci placed on each side of the center of the tube, the wetting bubble is stable to fluctuations in both directions. Of course, bubbles come in a range of sizes and this asymmetry between the two phases is more pronounced for smaller bubbles. On the average we can say that the nonwetting phase when it is immobilized will block more nodes than tubes. The wetting phase when immobilized will occupy comparatively more tubes than nodes. It is more effective to block nodes than tubes, and this explains why the nonwetting phase is more effective than the wetting phase in reducing the specific permeability of the other fluid. The result is that the pressure curve is steeper in the first region than in the third, and thus the first region is narrower than the third.

\subsection{Relationship between $F_{\rm nw}$ and $\Delta P$}
\label{sec:presanddfds}

This subsection is devoted to the study of the central range of saturation where both phases are mobilized. In this region we have defined the crossover point as the point where the fractional flow is equal to the saturation. On both sides of this point the phase whose saturation is higher gains in the sense that the fractional flow is higher than the saturation. In this region the fractional flow curves are roughly S-shaped. In particular, for the capillary numbers in Fig. \ref{fig:fflows}(c)-(d), the curvature changes approximately at the crossover point. Whether this is coincidental will be discussed later. We have determined the derivative of all the six curves. From these data we have estimated the turning points of zero curvature which are listed in Table \ref{thetablei}.

The differentiation has been done in the straightforward way, except at the boundaries of the central region. Outside the region the derivative is zero, and inside the region the curves are smooth. However, right at the boundary the fractional flow curves have, for the lower capillary numbers, a break where the derivative almost diverges. We have ignored these breaking points.

We find that the derivative has the same shape as the global pressure in the central region. This is shown in Fig. \ref{fig:globpres}, where the derivative is marked by `$\circ$'. By the same shape we mean that we for each capillary number can find two dimensionless constants $A\ $마nd $B\ $맍or which
\begin{equation}
  A(C_{\rm a}) \times \frac{dF_{\rm nw}}{dS_{\rm nw}} + B(C_{\rm a}) =
  \frac{\Delta P}{\Delta P_{\rm s}} .
\label{eq:scaleab}
\end{equation}
The quality of the overlap of the two sets of data is very good for the four highest capillary numbers, Fig. \ref{fig:globpres}(a)-(d). For the fifth, Fig. \ref{fig:globpres}(e) the quality is fair. For the lowest capillary number the points start to become spread out so much that a comparison of shapes is difficult. We have included the derivative here to get an estimate for the scaling coefficients $A\ $마nd $B$. The values of the scaling coefficients are listed in Table \ref{thetableii}. 

\begin{figure}
\centering
\includegraphics[width=7cm]{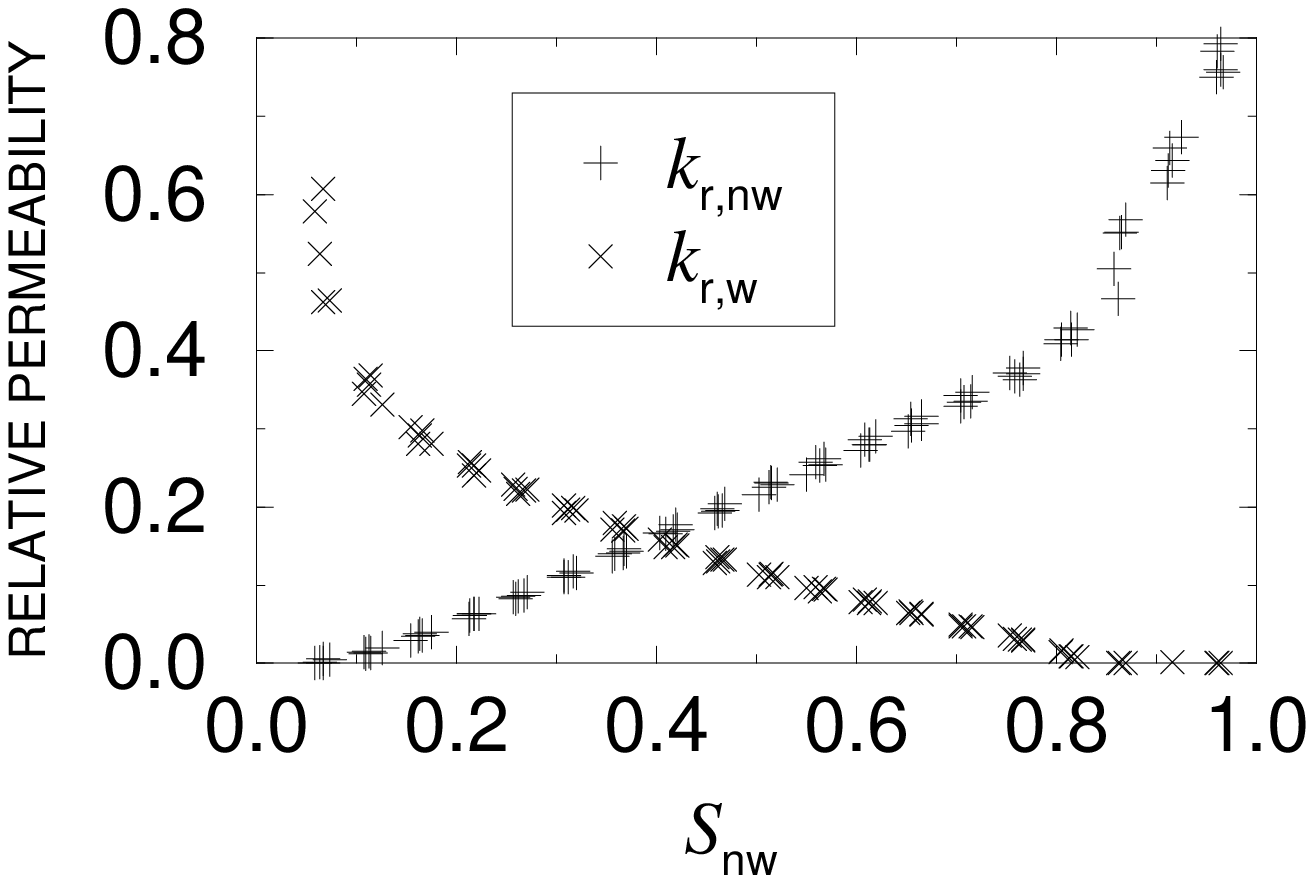}
\caption{The nonwetting and wetting relative permeability is shown as a function of nonwetting saturation. They are defined in Eqs. (\ref{eq:relnw}) and (\ref{eq:relw}). The capillary number is $C_{\rm a}=3.2\times 10^{-3}$, and these curves correspond directly to the fractional flow and pressure curves which are found in Fig. \ref{fig:fracpres}.}
\label{fig:relperm}
\end{figure}

In order to understand the meaning of Eq. (\ref{eq:scaleab}), it is useful to rewrite the expression in the terminology of mobilities and relative permeabilities\cite{D92,HRAFJH97}. The nonwetting relative permeability $k_{\rm r,nw}$ is defined by
\begin{equation}
  \frac{Q_{\rm nw}(S_{\rm nw})}{\Sigma} = 
  \frac{k_{\rm r,nw}(S_{\rm nw}) k}{\mu_{\rm nw}}
  \times \frac{\Delta P (S_{\rm nw})}{L},
\label{eq:qnwfrac}
\end{equation}
where the fraction
\begin{equation}
  M_{\rm nw}(S_{\rm nw}) =
  \frac{k_{\rm r,nw}(S_{\rm nw})}{\mu_{\rm nw}}
\label{eq:mobility1}
\end{equation}
is the nonwetting mobility. Here the constant $k\ $말s the specific permeability as it was defined in Eq. (\ref{eq:darcy}). Substituting for $k$, keeping in mind that both phases have the same viscosity $\mu$, and expressing $Q_{\rm nw}\ $말n terms of $F_{\rm nw}$, we obtain
\begin{equation}
  k_{\rm r,nw}(S_{\rm nw}) = F_{\rm nw}(S_{\rm nw})
  \times \frac{\Delta P_{\rm s}}{\Delta P(S_{\rm nw})}.
\label{eq:relnw}
\end{equation}
Likewise the result for wetting relative permeability is
\begin{equation}
  k_{\rm r,w}(S_{\rm nw}) = F_{\rm w}(S_{\rm nw})
  \times \frac{\Delta P_{\rm s}}{\Delta P(S_{\rm nw})},
\label{eq:relw}
\end{equation}
where the single phase pressure drop $\Delta P_{\rm s}$ is the same because we consider two fluids with equal viscosities. The relative permeabilities for the capillary number $C_{\rm a}=3.2\times 10^{-3}$ are shown in Fig. \ref{fig:relperm}. These curves are comparable to experimentally obtained relative permeability curves which are frequently used in the petroleum industry\cite{D92}.

Analogously to nonwetting mobility in Eq. (\ref{eq:mobility1}), we define wetting mobility as
\begin{equation}
  M_{\rm w}(S_{\rm nw}) =
  \frac{k_{\rm r,w}(S_{\rm nw})}{\mu_{\rm w}},
\end{equation}
and the total mobility as
\begin{equation}
  M(S_{\rm nw}) = M_{\rm nw}(S_{\rm nw}) + M_{\rm w}(S_{\rm nw}).
\end{equation}
In the simulations the total flux has been held constant. It can be expressed by using Eq. (\ref{eq:qnwfrac}), and its wetting counterpart, as
\begin{align}
  Q_{\rm tot} & = Q_{\rm nw} + Q_{\rm w} = 
  \Sigma (M_{\rm nw} + M_{\rm w}) k \times
  \frac{\Delta P}{L} \nonumber \\
  & = \frac{k\Sigma}{L} \times M(S_{\rm nw}) \Delta P(S_{\rm nw}),
\end{align}
which can be solved for $\Delta P$; in combination with (\ref{eq:darcy}) this is right hand side of Eq. (\ref{eq:scaleab}). Likewise the fractional flow can be expressed in terms of the mobilities,
\begin{equation}
  F_{\rm nw} = \frac{M_{\rm nw}}{M}.
\end{equation}
Thus the scaling relation in Eq. (\ref{eq:scaleab}) can be rewritten as
\begin{equation}
  A \times \frac{d}{dS_{\rm nw}} \left(
  \frac{M_{\rm nw}(S_{\rm nw})}{M(S_{\rm nw})} \right) + B
  = \frac{1}{\mu M(S_{\rm nw})},
\end{equation}
where $A\ $마nd $B$ are independent of the saturation. Differentiating once more, we find
\begin{equation}
  A \times \frac{d^2}{dS^2_{\rm nw}} \left(
  \frac{M_{\rm nw}(S_{\rm nw})}{M(S_{\rm nw})} \right)
  = -\frac{1}{\mu M^2}\times \frac{dM(S_{\rm nw})}{dS_{\rm nw}}.
\label{eq:scalemob}
\end{equation}
The new insight given by this equation is as follows. Generally the nonwetting and wetting mobility are considered as two independent functions of saturation. The result in Eq. (\ref{eq:scalemob}) shows that the two mobilities are related through an equation.

The dependence which we have found is so far restricted to the case where both phases have equal viscosity. Whether the result will be extendable to the case of two different viscosities in some form is an interesting question, but has been outside the scope of the present work. The validity is also restricted to capillary numbers in the range $3.2\times 10^{-4} < C_{\rm a} < 3.2\times 10^{-2}$. For higher capillary numbers there is little reason to expect any new interesting behavior since the asymptotic behavior of the fractional flow curve is the straight diagonal. Lower capillary numbers than this range are more interesting, also from a practical point of view since they may occur in reservoir conditions. Challenges in this region of parameter space are increased history dependence of the results, and considerably increased CPU time.

The validity of Eq. (\ref{eq:scaleab}) comes from visual inspection of the data collapse in Fig. \ref{fig:globpres}. Visually it seems that for each capillary number, the two curves have their maximum value at the same saturation. We wish to discuss whether these maximum points are coincidental with the crossover points on the fractional flow curves. Estimates for all three points for each capillary number are listed in Table \ref{thetablei}. The error estimates come from the data analysis, and for the turning point precision is lost in the differentiation process. The crossover points are sensitive to systematic errors, in particular for high capillary numbers. The reason is that the fractional flow curve becomes increasingly parallel to diagonal for increasing capillary numbers. A small vertical shift of the fractional flow curve will give a large shift in the crossing point. From the values in the table we observe that the three points coincide for the four lowest capillary numbers within the error bars given. That is, the turning point for the two lowest capillary numbers are very uncertain and not listed respectively. For the two highest capillary numbers the crossover point is not equal to the maximum pressure within the error bars, but again one may speculate in the possibility of systematic errors in these two points.

In conclusion, we cannot say whether the crossover point is related with the maximum point. However, it is an interesting question because it is suggestive to look for effective theoretical descriptions starting at a point having this supposedly nice behavior. Experimental work on this problem will provide means of checking both Eq. (\ref{eq:scaleab}) and the possible relation between the crossover and the maximum pressure point.

\begin{table}
\caption{The table lists the crossover points of the fractional flow curves(crossover), the maximum points of the pressure curves(maximum $\Delta P$), and the turning points(turn) of the fractional flow curves for six different capillary numbers. The points are all saturation values, which are dimensionless numbers in the range from zero to one. The three points seem to coincide for the four lowest capillary numbers. The crossover point differs from the other two for the two highest capillary numbers.}
\label{thetablei}
\begin{tabular}{llll}
$C_{\rm a}$ & crossover & maximum $\Delta P$ & turn \\
\hline
$3.2\times 10^{-2}$ & 0.13(2) & 0.19(4) & 0.22(4) \\
$1.0\times 10^{-2}$ & 0.18(2) & 0.22(5) & 0.23(6) \\
$3.2\times 10^{-3}$ & 0.25(2) & 0.28(5) & 0.25(7) \\
$1.0\times 10^{-3}$ & 0.33(2) & 0.31(4) & 0.34(4) \\
$3.2\times 10^{-4}$ & 0.40(2) & 0.38(5) & 0.31(8) \\
$1.0\times 10^{-4}$ & 0.48(3) & 0.45(6) & --- \\
\end{tabular}
\end{table}

\begin{figure}
\centering
\includegraphics[width=7cm]{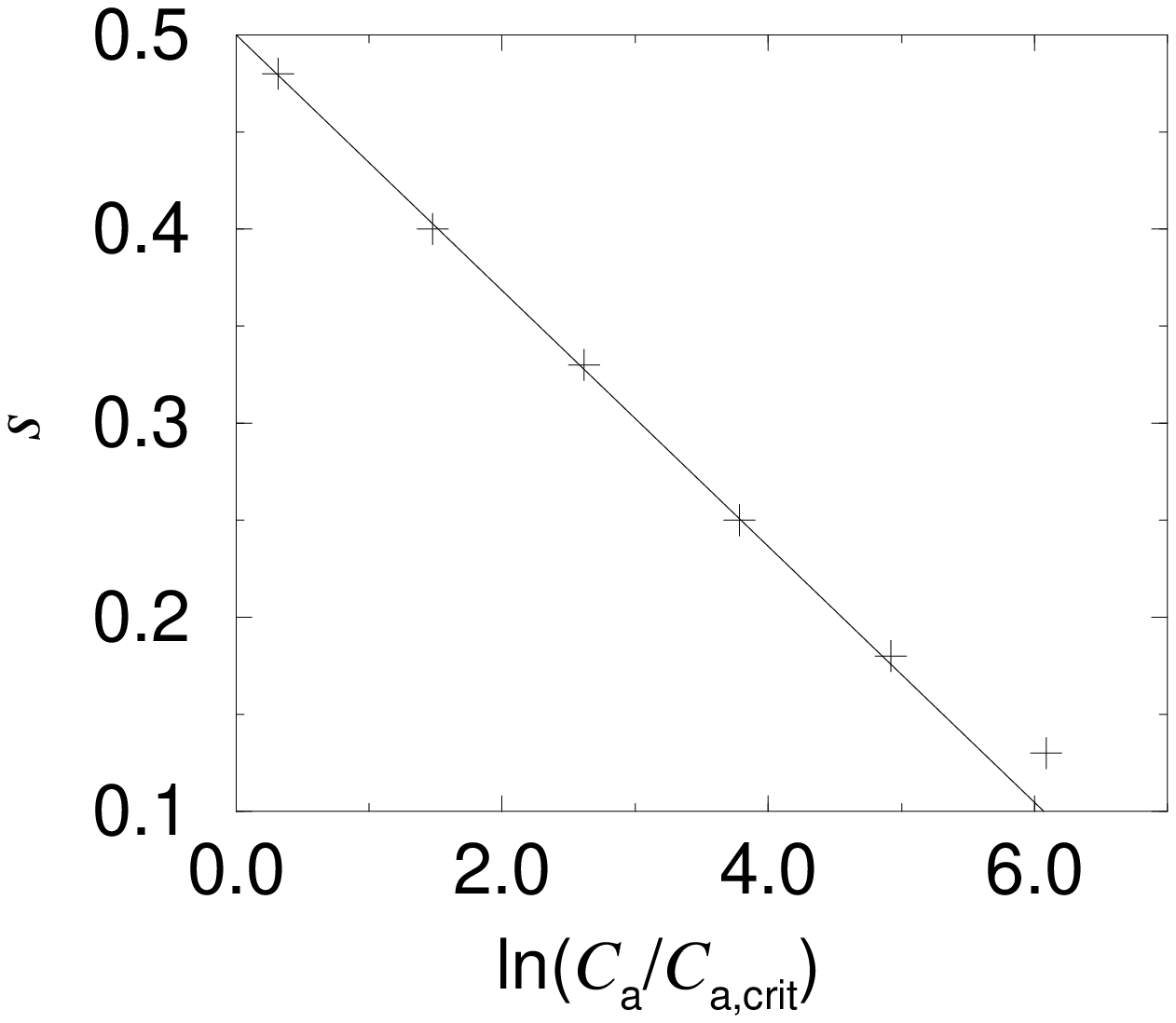}
\caption{The crossover points $s$ from Table \ref{thetablei} are shown as a function of the logarithm of the capillary number. The straight line fit is made on the basis of the four lowest capillary numbers, which are closest to the critical capillary number.}
\label{fig:cross}
\end{figure}

It is interesting to look at how the crossover point and scaling factors vary with the capillary number. We argued that the crossover point should approach the percolation threshold $s_{\rm c}$ in the limit of small capillary numbers. This limit turns out to be not zero but a lower critical capillary number $C_{\rm a,crit}$. We denote the crossover point, the crossover saturation, $s$. The crossover point as a function of the logarithm of the capillary number is shown in Fig. \ref{fig:cross}. That is, the capillary number is normalized with respect to the critical capillary number which is determined from fitting the following relation to the data points,
\begin{equation}
s=s_{\rm c} - a \ln{\left( \frac{C_{\rm a}}{C_{\rm a,crit}} \right)}.
\end{equation}
Here $s_{\rm c}$ is known so that the value of $a=0.066$ is the slope of a straight line fit, and the value of $C_{\rm a,crit}=7.3\times 10^{-5}$ is the one which makes the line meet the $s$-axis at $s_{\rm c}=0.5$.

The physical idea behind having a critical capillary number is that at some point the viscous forces of the flowing fluids and the pressure gradient will become so small that they cannot mobilize any more interfaces. In that situation one of the phases will have a continuous pathway to flow in. The other phase is pinned to its current locations. Therefore any further decrease in the capillary number below the critical value, will not add anything to the picture; one phase flows.

The scaling coefficients $A$ and $B$ from Eq. (\ref{eq:scaleab}) are listed in Table \ref{thetableii}. The dimension of these coefficients are the same as for pressure. The immediate tentative interpretation of the meaning of the two are as follows. The constant $B$ is the threshold pressure which is applied right at the borders of the central region of the fractional flow curves, the pressure where there is onset of mobility of both phases. The factor $A\ $말s the one that must be multiplied to the derivative of the fractional flow curve in order to obtain the pressure jump from the onset level $B$ to the actual pressure at a given saturation in the central region.

\begin{figure}
\centering
\includegraphics[width=7cm]{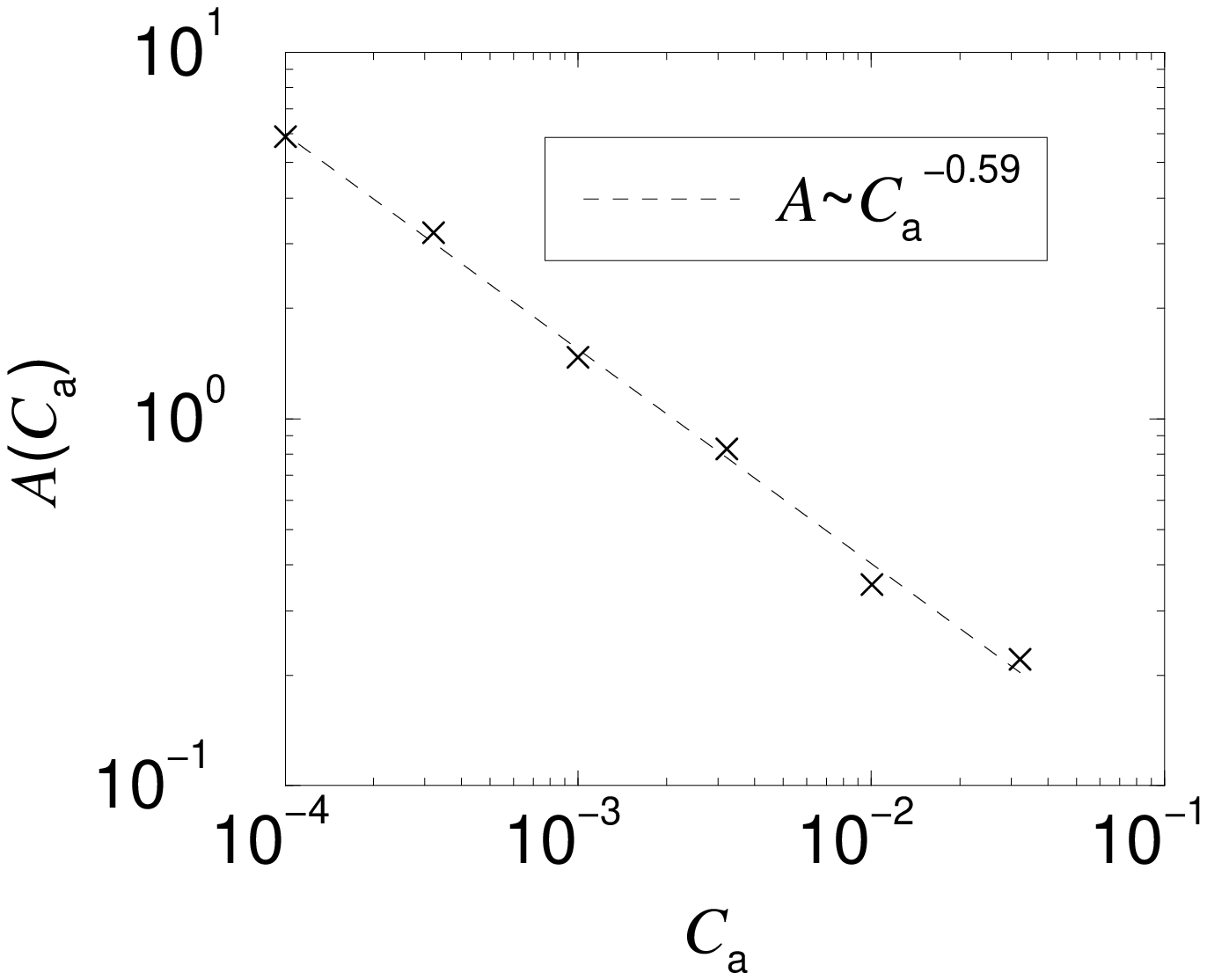}
\caption{The scaling factor $A$ from Eq. (\ref{eq:scaleab}). Data are taken from Table \ref{thetableii}.}
\label{fig:factor}
\end{figure}

\begin{figure}
\centering
\includegraphics[width=7cm]{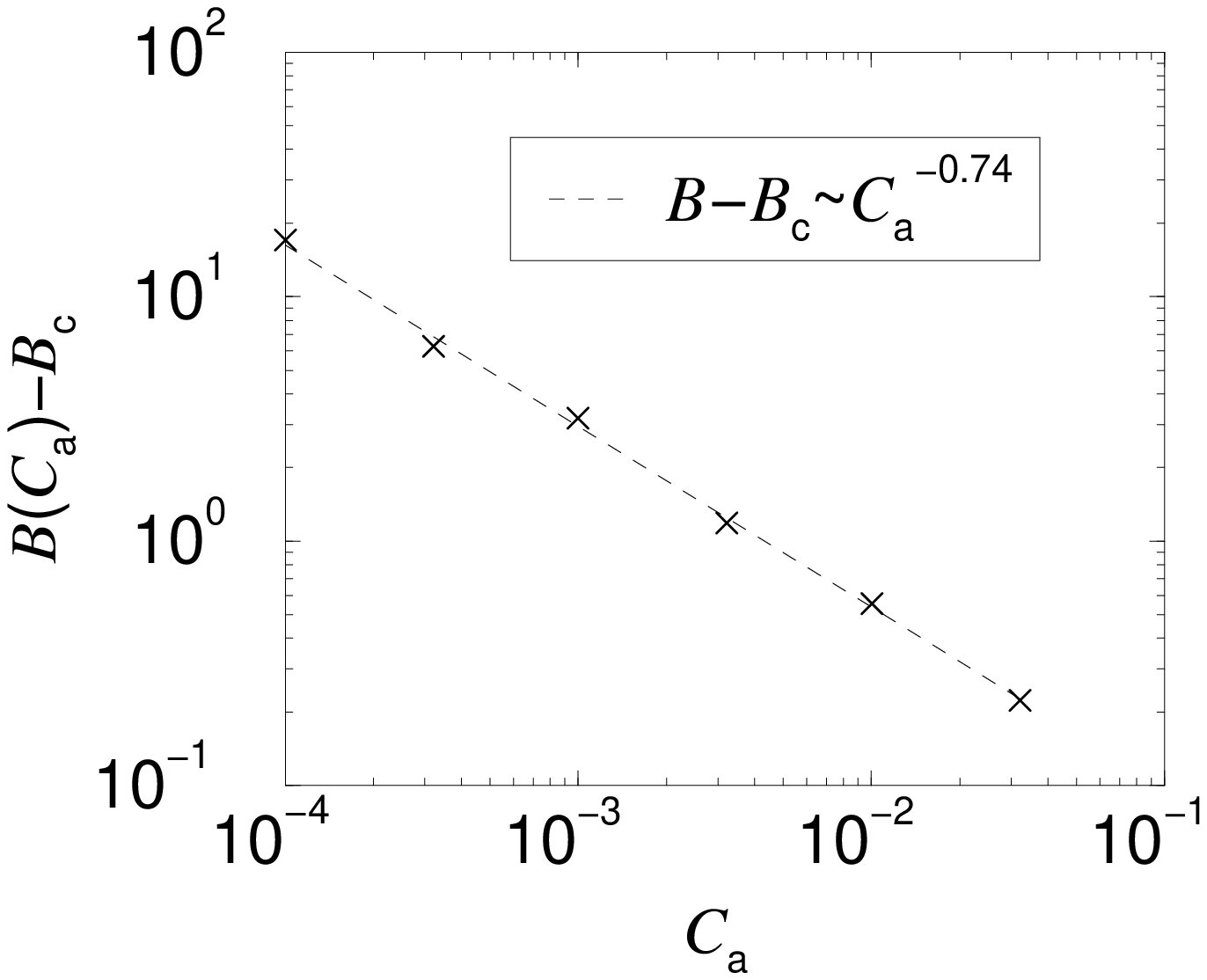}
\caption{The scaling factor $B$ from Eq. (\ref{eq:scaleab}). Data are taken from Table \ref{thetableii}. The value of $B_{\rm c}$ is the one which gives the best power law fit; $B_{\rm c} = 0.65$.}
\label{fig:modconst}
\end{figure}

It is clear that the pressure is an increasing function of the capillary number. Therefore it is natural that $A$ and $B$ are functions of the capillary number. The functional forms are presented in Figs. \ref{fig:factor} and \ref{fig:modconst}. We can say that the explanation of these two power laws is related to the width of the central region. Also the extent to which interfaces are put in motion plays a role. Both these quantities depend on the capillary number. We have so far not been able to construct sound explanations for these power laws, and present them as observations only.

\begin{table}
\caption{The scaling factors $A$ and $B$ from Eq. (\ref{eq:scaleab}) are shown for the six capillary numbers. They are used in the scaling of the derivative of the fractional flow curves which is shown in Fig. \ref{fig:globpres}.}
\label{thetableii}
\begin{tabular}{lll}
$C_{\rm a}$ & $A(C_{\rm a})$ & $B(C_{\rm a})$ \\
\hline
$3.2\times 10^{-2}$ & 0.22 & 0.87  \\
$1.0\times 10^{-2}$ & 0.35 & 1.21  \\
$3.2\times 10^{-3}$ & 0.83 & 1.84  \\
$1.0\times 10^{-3}$ & 1.47 & 3.83  \\
$3.2\times 10^{-4}$ & 3.22 & 6.90  \\
$1.0\times 10^{-4}$ & 5.89 & 17.6  \\
\end{tabular}
\end{table}

\section{Discussion}
\label{sec:discussion}
The problem of two-phase flow in porous media is complex. We attack the problem with a network simulator on pore level. The model is coarse grained on the level of the internal structure of the pores. The current use of the model is generic, but specific use is possible with few modifications. Useful average properties of porous media can be obtained which may be used as parameters on larger reservoir scale simulations.

In this paper we present results for 2D square networks of tubes. This topology is chosen for convenience. It is possible to extend the work to 3D and choose both regular and irregular connectivity. The choice of a particular topology is one aspect of making the model specific to a given porous medium. Further, average tube radius and the the width of the tube radius distribution are important parameters that should be adjusted for the same purpose. Here, we have used a distribution which corresponds to experimental sizes in Hele-Shaw cells with glass beads which have been used experimentally\cite{FMSH97}. Further work and comparisons with these experiments may provide insight into how the model can be callibrated in order to become quantitavely precise.

The most important characteristic of the simulations in this paper, which sets them apart from other simulations, is the biperiodic boundary conditions. This makes the system closed with a fixed saturation of each of the two phases. For six different capillary numbers we run the systems until they reach a steady state where the flow is characterized by complex bubble dynamics. The notion of imbibition and drainage are not adequate to describe this situation, which would also be the situation deep inside reservoirs. We find average flow properties as a function of the saturation. These properties are the fractional flow and the global pressure drop, from which in turn one can calculate relative permeabilities and mobilities.

In particular for the case of two phases having equal viscosity, we find that the derivative of the fractional flow is related to the global pressure drop, see Eq. (\ref{eq:scaleab}). This relation ties together these two properties which generally are believed to be independent. We have shown how this statement can be transformed into the language of relative permeabilities and mobilities, which as a consequence are not independent but must obey Eq. (\ref{eq:scalemob}).

An important note here is that this result so far has only been established by numerical work. It is very interesting to get an experimental verification of this result. Hopefully, after an experimental verification, this equation can be of assistance in the measurement of two-phase flow properties. In the experimental situation it is difficult to measure all variables precisely. The saturation can e.g. be reconstructed from the pressure and fractional flow relationship and Eq. (\ref{eq:scaleab}).

\section*{Acknowledgements}
H.A.K. acknowledges support from VISTA, a collaboration between Statoil and The Norwegian Acad. of Science and Letters. We acknowledge support from the CNRS through a PICS grant and through G. G. Batrouni and INLN for hospitality during part of this work. We thank Eyvind Aker and Knut J{\o}rgen M{\aa}l{\o}y for valuable comments.


%
%

%
%


\begin{references}
\bibitem[*]{emailhak} Electronic address: Henning.Knudsen@phys.ntnu.no
\bibitem[**]{emailah} Electronic address: Alex.Hansen@phys.ntnu.no
\bibitem{CW85} J.-D. Chen and D. Wilkinson, Phys. Rev. Lett. {\bf 55}, 1892 (1985).
\bibitem{MFJ85} K. J. M{\aa}l{\o}y, J. Feder, and T. J{\o}ssang, Phys. Rev. Lett. {\bf 55}, 2688 (1985).
\bibitem{LTZ88} R. Lenormand, E. Touboul, and C. Zarcone, J. Fluid Mech. {\bf 189}, 165 (1988).
\bibitem{KL85} J. Koplik and T. J. Lasseter, SPE J{\bf 2}, 89 (1985).
\bibitem{BK90} M. Blunt and P. King, Phys. Rev. A {\bf 42}, 4780 (1990).
\bibitem{R90} D. H. Rothman, J. Geophys. Res. {\bf 95}, 8663 (1990).
\bibitem{CP91} G. N. Constantinides and A. C. Payatakes, J. Coll. Int. Sci. {\bf 141}, 486 (1991).
\bibitem{CP96} G. N. Constantinides and A. C. Payatakes, AIChE J. {\bf 42}, 369 (1996).
\bibitem{WS81} T. A. Witten and L. M. Sander, Phys. Rev. Lett. {\bf 47}, 1400 (1981).
\bibitem{WW83} D. Wilkinson and J. F. Willemsen, J. Phys. A {\bf 16}, 3365 (1983).
\bibitem{P84} L. Paterson, : 1984, Phys. Rev. Lett. {\bf 52}, 1621 (1984).
\bibitem{BC98} P. Binning and M. A. Celia, Advances in Water Resources {\bf 22}, 461 (1998), and references herein.
\bibitem{KAH00} H. A. Knudsen, E. Aker, and A. Hansen., To appear in Trans. in Por. Med., cond-mat/0008014.
\bibitem{RK88} D. H. Rothman and J. M. Keller, J. Stat. Phys. {\bf 52}, 1119 (1988).
\bibitem{S95} M. Sahimi, \emph{Flow and Transport in Porous Media and Fractured Rock} (VCH, New York, 1995)
\bibitem{P77} D. W. Peaceman, \emph{Fundamentals of numerical reservoir simulations} (Elsevier, Amsterdam, 1977).
\bibitem{AMHB98} E. Aker, K. J. M{\aa}l{\o}y, A. Hansen, and G. G. Batrouni, Trans in Por. Med. {\bf 32}, 163 (1998)
\bibitem{AMH98} E. Aker, K. J. M{\aa}l{\o}y, and A. Hansen, Phys. Rev. E {\bf 58} (1998)
\bibitem{AP95a} D. G. Avraam and A. C. Payatakes, J. Fluid Mech. {\bf 293}, 207 (1995).
\bibitem{AP95b} D. G. Avraam and A. C. Payatakes, Trans. in Porous Media {\bf 20}, 135 (1995).
\bibitem{AP99} D. G. Avraam and A. C. Payatakes, Ind. Eng. Chem. Res. {\bf 38}, 778 (1999).
\bibitem{Roux} S. Roux, unpublished.
\bibitem{W21} E. W. Washburn, Phys. Rev. {\bf 17}, 273 (1921)
\bibitem{VP01} M. S. Valavanides and A. C. Payatakes, Adv. in Wat. Res. {\bf 24}, 385 (2001).
\bibitem{D92} F. A. L. Dullien, \emph{Porous Media: Fluid Transport and Pore Structure} (Academic Press, San Diego, 1992).
\bibitem{HRAFJH97} A. Hansen, S. Roux, A. Aharony, J. Feder, T. J{\o}ssang, and H. H. Hardy, Trans. in Por. Med. {\bf 29}, 247 (1997).
\bibitem{SA92} D. Stauffer and A. Aharony, \emph{Introduction to Percolation Theory} (Taylor \& Francis, London, 1994).
\bibitem{FMSH97} O. I. Frette, K. J. M{\aa}l{\o}y, J. Schmittbuhl, and A. Hansen, Phys. Rev. E {\bf 55}, 2969 (1997). 
\end{references}
\end{document}